# A Fitch-Based Wind Farm Parameterization for CM1: Implementation, Validation, and Impacts on Convection Initiation

Jianxiang Li and Zhiyong Meng

*Peking University, Beijing, China*

**Key Points:**

- A Fitch-based wind farm scheme is implemented in Cloud Model 1 and validated against the Weather Research and Forecasting model.
- Under stable conditions, organized near-surface wind and temperature perturbations extend well beyond the wind farm.
- Large turbines can promote convection initiation tens of kilometers downstream through gravity waves and boundary layer warming.

## Abstract

We implement a Fitch-based wind farm parameterization in Cloud Model 1 (CM1), representing turbines as elevated momentum sinks and turbulent kinetic energy (TKE) sources within the rotor-swept layer, with optimizations for computational efficiency and parallel execution. Direct comparison with the Weather Research and Forecasting (WRF) model shows close agreement in the magnitude and spatial structure of wind-speed deficits and TKE enhancement. Diurnal simulations also capture the known dependence of boundary layer responses on atmospheric stability, further supporting the consistency of the implementation. Under stable conditions, the wind farm produces coherent spatial patterns in near-surface wind speed and temperature that extend well beyond the farm. In idealized simulations under conditions favorable for convection, large turbines promote convection initiation tens of kilometers downstream, earlier than in the no-farm simulation. Analysis identifies turbulent entrainment of warmer air across the boundary layer top as the dominant source of an elongated downstream warm band, with a smaller contribution from enhanced surface sensible heat flux. Secondary circulations develop across this warm band and provide sustained low-level lifting. A gravity wave generated by the wind farm supplies the initial vertical perturbation, which intensifies as it is advected downstream within an ascending branch of these circulations and develops into a localized deep convective updraft.

## Plain Language Summary

We added a computationally efficient representation of wind farms to Cloud Model 1 to simulate how turbines slow winds and generate turbulence. Comparisons with another atmospheric model show close agreement in both the strength and spatial patterns of wind-speed reductions and increased turbulence. Simulations spanning day and night also reproduce known differences in how wind farms affect the lower atmosphere. Under stable atmospheric conditions, the wind farm produces organized patterns of near-surface wind speed and temperature changes that extend well beyond the farm. In idealized simulations under favorable conditions, large turbines promote the development of rising air and convective clouds tens of kilometers downstream, earlier than without a wind farm. An elongated warm band forms downstream, mainly because turbulence mixes air from above into the lower atmosphere. Increased heat transfer from the ground plays a smaller role. Circulations develop across the warm band and sustain upward air motion. A wave generated by the wind farm provides the initial lift. As the disturbance travels downstream within the rising branch of these circulations, it strengthens into a strong, localized updraft.

# 1. Introduction

Severe convective storms can produce tornadoes, large hail, and damaging winds, causing substantial loss of life and property within a short time. Their small scales and rapid, nonlinear evolution make their internal structure difficult to observe and their initiation and development difficult to predict. Idealized numerical experiments provide a controlled setting in which individual thermodynamic, dynamical, and microphysical influences on storm initiation and evolution can be isolated.

Global installed wind capacity has grown at a compound annual growth rate of over 10% during the past 25 years, reaching 1299 GW in 2025, with continued rapid expansion projected (GWEC, 2026). Turbine rated power has also increased, accompanied by larger hub heights and rotor diameters (GWEC, 2025). As wind energy expands, wind farms are emerging as a distinct type of anthropogenic landscape, with turbines acting as elevated momentum sinks and turbulence sources. Their expansion and increasing turbine size motivate examination of how this forcing modifies planetary boundary layer (PBL) processes and mesoscale atmospheric phenomena, including the thermodynamic and dynamic conditions relevant to convection.

Observations and numerical simulations show that wind farms can alter PBL structure. At night, turbine-induced turbulence enhances vertical mixing and warms the near-surface atmosphere through turbulent heat flux divergence and three-dimensional temperature advection (Xia et al., 2019), strengthening downward surface sensible heat flux and raising surface temperatures. Most field, satellite, and numerical studies report nocturnal warming, although results vary with terrain and season; daytime effects are generally insignificant or show slight cooling (Baidya Roy & Traiteur, 2010; Jia et al., 2024; Jiang et al., 2025; Zhou et al., 2022; Zhou et al., 2012). Wind-speed deficits can extend tens of kilometers downstream under stable conditions and recover more rapidly under unstable conditions (Englberger & Dörnbrack, 2017; Porté-Agel et al., 2020). Near-surface wind speeds tend to decrease during the day and increase at night (Fitch, Lundquist, et al., 2013; Rajewski et al., 2013), although this diurnal response is not always present (Bodini et al., 2021). Wind farms also tend to enhance surface evaporation and reduce soil moisture (Jia et al., 2024), with different responses possible when vegetation or irrigation is included (Liu et al., 2023; Su et al., 2025).

Wind farms can weaken and elevate boundary-layer low-level jets and reduce their frequency (Quint et al., 2025; Zhou et al., 2022). Large wind farms can also excite gravity waves in the capping inversion and the free atmosphere above, generating pressure perturbations that modify the wind field within and upstream of the farm (Allaerts & Meyers, 2017; Smith, 2010), and slow the propagation of cold-pool outflow boundaries (Tomaszewski & Lundquist, 2021).

Simulations and observations indicate that large wind farms can alter regional precipitation, with reported responses including upstream increases, downstream decreases, and greater warm-season rainfall (Al Fahel & Archer, 2020; Dong et al., 2026; Fiedler & Bukovsky, 2011). These studies primarily address regional or time-averaged precipitation, which does not establish whether or how wind farms affect the initiation and development of individual convective systems. Wind farm effects on convection initiation (CI), in particular, have received very limited attention. Meng et al. (2026) presented a case study showing that a desert wind farm could modify a boundary layer convergence line and accelerate CI through gravity-wave and cloud–radiative feedbacks. However, the generality of these effects and their dependence on atmospheric conditions and wind farm characteristics remain unclear, motivating further investigation through idealized simulations.

Several approaches exist for representing wind farms in numerical models. Implicit parameterizations increase surface roughness length over wind farm areas (Fischereit et al., 2021), an approach criticized as inappropriate for representing wind farms and their impacts (Fitch, Olson, et al., 2013). The Fitch scheme (Fitch et al., 2012) represents turbines as elevated momentum sinks and turbulent kinetic energy (TKE) sources. It is widely used in WRF (Fischereit et al., 2021) and has been evaluated against field observations and LES simulations (Agarwal et al., 2026; García-Santiago et al., 2024; Peña et al., 2022), with subsequent developments addressing rotor-scale wind variability, axial induction, turbine layout, and subgrid wake effects (Ali et al., 2026; Chang et al., 2026; Du et al., 2025; Du et al., 2026; Ma et al., 2022; Redfern et al., 2019; Vollmer et al., 2024). The Volker scheme (Volker et al., 2015), an explicit wake parameterization based on wake expansion theory, adds momentum sinks without explicitly injecting TKE. At the LES scale, wind turbines are represented as actuator disk models, actuator line models or actuator surface models (Li et al., 2022), resolving individual wakes at a computational cost that limits domain size and simulation duration.

Cloud Model 1 (CM1; Bryan & Fritsch, 2002) is a three-dimensional, compressible, nonhydrostatic model designed for idealized studies of atmospheric convection and mesoscale dynamics. Its open-source code, computational efficiency, and flexible configuration support high-resolution simulations and systematic sensitivity experiments. Control over the initial environment and physical processes allows individual influences on storm structure and evolution to be isolated. These capabilities make CM1 a useful numerical platform for investigating the thermodynamic, dynamical, and microphysical mechanisms of convection initiation and organization. Before this work, however, CM1 lacked a dedicated wind farm parameterization. For the kilometer-scale simulations considered here, we adopt the Fitch scheme because it is the most widely used wind farm parameterization in mesoscale modeling (Fischereit et al., 2021).

Here, we implement a Fitch-based wind farm parameterization in CM1 and optimize it for computational efficiency and parallel execution. We evaluate the implementation against WRF and examine the diurnal boundary-layer response. We also analyze the spatial organization of downstream wind and temperature perturbations under stable conditions. In a separate convection-favorable environment, we test the sensitivity of CI to turbine configuration and analyze its mechanism.

# 2. Model and Parameterization Description

## 2.1 Wind Farm Parameterization

### 2.1.1 The Fitch parameterization

The parameterization is adapted from the Fitch wind farm parameterization in WRF. Each wind turbine is treated as an elevated momentum sink and a source of TKE within the grid cells intersecting its rotor disk. Let $U_k = (u_k, v_k)$ denote the horizontal velocity components in layer $k$, and $z_k$ *and* $z_{k+1}$ the layer-interface heights; $t$ denotes time. At each horizontal grid $(i, j)$, the tendency equations for wind speed and TKE are driven by the wind speed valid at each vertical level $k$:

$$\frac{\partial \mathrm{u_k}}{\partial \mathrm{t}} = -\sum_{T} \frac{\frac{1}{2} N_T C_t |\boldsymbol{U_k}| u_k A_k}{z_{k+1} - z_k} \tag{1}$$

$$\frac{\partial \mathrm{v_k}}{\partial \mathrm{t}} = -\sum_{T} \frac{\frac{1}{2} N_T C_t |\boldsymbol{U_k}| v_k A_k}{z_{k+1} - z_k} \tag{2}$$

$$\frac{\partial \mathrm{TKE_k}}{\partial \mathrm{t}} = \sum_{T} \frac{\frac{1}{2} N_T C_{TKE} |\boldsymbol{U_k}|^3 A_k}{z_{k+1} - z_k} \tag{3}$$

Where $T$ represents the category of turbines, $N_T$ is the number density of T-type turbines per square meter in the horizontal grid cell, $A_k$ is the cross-sectional swept area at layer $k$, $C_\mathrm{t}$ is the thrust coefficient, $C_\mathrm{p}$ is the power coefficient, and the TKE source coefficient is given by $C_{TKE} = \alpha(C_t - C_p)$, with $\alpha = 0.25$ the fraction of extracted kinetic energy (Archer et al., 2020). $C_p$ and $C_t$ are functions of $|\boldsymbol{U_H}|$, which is the wind speed magnitude at the turbine hub height. Note that $C_t$, $C_p$, $C_{TKE}$ and $A_k$ also depend on the turbine category T. With $\rho$ the air density and $A_T$ the full rotor-swept area of a T-type turbine, power production per unit horizontal area is defined as:

$$P = \sum_{T} \frac{1}{2} C_p \rho |\boldsymbol{U_H}|^3 A_T N_T \tag{4}$$

To ensure these multilayer estimates conserve energy and match total wind energy production derived from the hub-height wind speed, all tendency equations (1)–(3) are multiplied by a normalization factor (ec) for each turbine category T on the right-hand side:

$$ec = \frac{\frac{1}{2}C_p\rho_a|\boldsymbol{U_H}|^3 A_T N_T}{\sum_{k=turbine\ bottom}^{turbine\ top}\frac{1}{2}C_p\rho_a|\boldsymbol{U_k}|^3 A_k N_T} \tag{5}$$

Turbine specifications for the 2 MW case come from the WRF repository; those for 5, 10, 15, and 22 MW are from the NREL FLORIS library. The hub height and rotor diameter of wind turbines are shown in Figure 1.

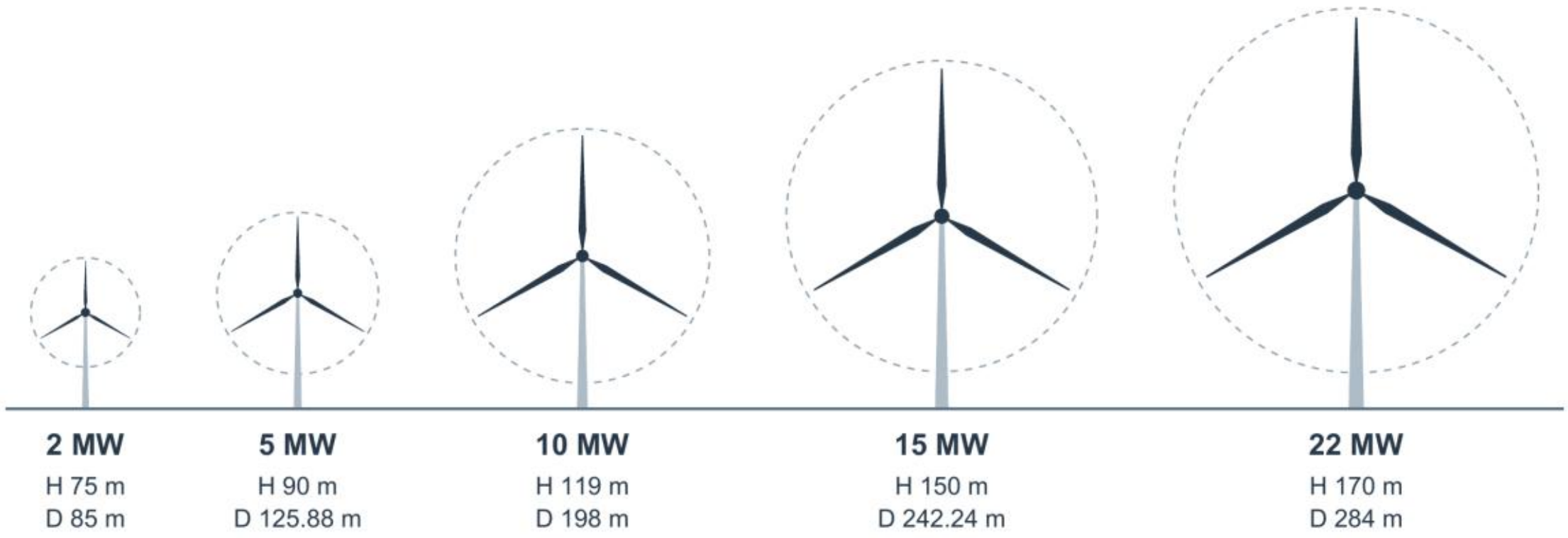

**Figure 1** Hub height (H) and rotor diameter (D) of wind turbines with different rated powers

### 2.1.2 Technical Implementation and Optimization

Beyond the basic WRF port, several technical enhancements improve efficiency and compatibility. Geometry-dependent parameters are precomputed once during initialization, avoiding redundant calculations at every time step. Data storage is restructured from per-turbine to per-turbine-type, cutting memory usage for large wind farms with many identical turbines. Turbine density calculations are generalized to support non-square grid cells. Native OpenMP (shared-memory) and MPI (distributed-memory) parallelism is provided. Comprehensive input validation prevents silent failures from malformed input, and computational loops are tuned to increase CPU cache hit rates. In addition to module_wind_fitch.F, several other files in the CM1 repository were modified, mainly to define variables and pass parameters.

## 2.2 CM1 and WRF configuration

To validate the scheme and investigate its impact on convection initiation, we conducted three sets of experiments, as will be discussed in Section 2.3. Except for the WRF control experiment in Experiment 1, all experiments were performed using CM1. In Experiment 1, the WRF simulations adopted the same configurations as CM1 to ensure a fair comparison.

All experiments adopt the Klemp-Wilhelmson compressible vertically implicit time-splitting integration scheme with a fixed time step of 3 seconds. To suppress 2Δx waves (Δx is horizontal grid spacing) and gray zone numerical artifacts, a 5th-order Weighted Essentially Non-Oscillatory (WENO) advection scheme for both scalar variables and velocity fields, a horizontal Smagorinsky turbulence parameterization scheme with coefficient $C_s = 0.25$ and a monotonic 6th-order artificial diffusion scheme with a coefficient of 0.12 are implemented.

All experiments use 1 km horizontal grid spacing. The vertical grid configuration and the damping layer differ between experiments. In Experiments 1 and 2, the model top is at 12.6 km, and a Rayleigh damping layer is applied from 10 km to the model top. In Experiment 3, the model top is raised to 21.6 km, and the damping layer starts at 15 km. The Rayleigh damping uses an e-folding time of 300 s in all experiments.

For all experiments, the vertical resolution is 12 m in the lowest 360 m and is then uniformly stretched to 600 m at 12.6 km. A uniform 600 m resolution extends from 12.6 km to the model top, except in Experiments 1 and 2, where the model top is set at 12.6 km. Since WRF uses a hybrid vertical coordinate, we adjust the eta levels to approximately match the vertical grid stretching in CM1.

The top boundary uses a free-slip condition, the bottom boundary uses a half-slip condition, and the lateral boundaries use periodic conditions. The initial temperatures at the surface and in the deep soil are set to 300 K. The boundary layer uses the MYNN Level 2.5 eddy-diffusivity mass-flux (EDMF) scheme with TKE advection option activated, the surface layer uses the MYNN scheme, and the land surface uses a slab scheme as CM1 does not include a more comprehensive land surface scheme. Experiments 2 and 3 use the RRTMG radiation scheme with a reference latitude of 45°N and the Morrison double-moment microphysics scheme, while these are turned off in Experiment 1. The radiation call interval is 300 s by default, but in Experiment 3 we set it to 60 s to simulate the interaction between clouds and radiation more accurately.

The Coriolis force (with Coriolis parameter $f = 0.0001$ $s^{-1}$) is activated in Experiment 1 and Experiment 2, where an additional horizontal pressure gradient corresponding to a geostrophic wind of 10 m/s in the zonal (u) direction is imposed to sustain a background wind field. For Experiment 3, the Coriolis force is disabled to eliminate unnecessary complexity.

All experiments were initialized with random potential temperature perturbations of 0.25 K at all vertical levels to break horizontal homogeneity and produce realistic boundary layer structures.

The simulated wind farm covers a 10 km × 10 km horizontal domain, consisting of 100 wind turbines with one turbine mounted at each horizontal grid point. The geometric center of the wind farm is located 50 km from the left boundary of the computational domain and is centered in the y direction.

## 2.3 Experimental Design

### 2.3.1 Experiment 1: WRF–CM1 Intercomparison

For all simulations in this experiment, we prescribed the USGS "Barren or Sparsely Vegetated" land surface category, with surface emissivity and moisture availability set to zero.

Referring to Fitch et al. (2012), we first conducted a precursor simulation in CM1 with the wind farm turned off to generate the initial sounding. This precursor run used a uniform u-wind of 10 m/s, a constant potential temperature of 300 K below 960 m, a lapse rate of 10 K/km aloft, and no moisture. The simulation was run for 10 days over a small 8 km × 8 km domain, and horizontal averaging was applied to produce the one-dimensional initial profile.

The formal simulations consisted of four experiments: CM1-WF and CM1-NWF (wind farm on and off in CM1), and WRF-WF and WRF-NWF (wind farm on and off in WRF). The formal runs were performed for 10 hours using a 400 km × 400 km domain. In the WF configurations, a wind farm consisting of 2-MW turbines was activated at the start of the simulation. The analysis focuses on a single time step at 10 h.

### 2.3.2 Experiment 2: Diurnal Cycle of Wind Farm Impacts

In Experiment 2, all simulations used the USGS "Barren or Sparsely Vegetated" land surface category with moisture availability set to 0.0001. This extremely dry setup was chosen to suppress moisture accumulation and avoid cloud and precipitation development during the long simulations. Therefore, moisture-related results should be interpreted qualitatively rather than quantitatively.

A precursor simulation was conducted on an 8 km × 8 km domain with the wind farm scheme turned off. This precursor run was initialized with the dry sounding same as that of the formal simulation in Experiment 1 and was integrated for 20 days (starting at 18:00 LT on 1 July 2025) to reach approximate radiative equilibrium; water vapor equilibrium cannot be attained. The horizontally averaged state from this run provided the initial sounding for the formal simulations.

The formal simulations consist of three groups: no wind farm (NWF), 2 MW turbines, and 22 MW turbines. They are run for 54 h over a 600 km × 200 km domain, starting at 18:00 LT on 21 July 2025. Analyses are performed after a 6-h spin-up, covering the period from 00:00 LT 22 July to 00:00 LT 24 July 2025.

### 2.3.3 Experiment 3: Convection Initiation

The initial atmospheric state is prescribed using the Weisman–Klemp thermodynamic and moisture profile (Weisman & Klemp, 1982), which has a near-surface water vapor mixing ratio maximum of 14 g/kg, along with a uniform westerly wind of 15 m/s. The land surface is set to the USGS "Wooded Wetland" category, and all simulations are initialized at 06:00 LT on 22 July 2025. The experiments are conducted for 6 h over a 300 km × 100 km domain. Six configurations are considered: a baseline case without a wind farm (NWF) and five wind farm cases with turbine rated powers of 2, 5, 10, 15, and 22 MW.

To avoid spurious transients, the wind farm is ramped up linearly between 1.5 h and 2.5 h by multiplying the right-hand side of the governing equations (1)–(3) by the dimensionless ramp factor

$$fac = \frac{t - t_{start}}{t_{ramp}} \tag{6}$$

for $t_{start} < t \le t_{start} + t_{ramp}$ with $t_{start} = 1.5\ h$ and $t_{ramp} = 1.0\ h$. Here $t$ is simulation time, and the subscripts "start" and "ramp" denote the ramp start time and duration, respectively. The wind farm is not activated at the beginning of the simulation because boundary-layer winds decrease during model spin-up, so applying turbine drag to the initially stronger flow would produce unrealistically large disturbances. The forcing is then increased gradually rather than switched on instantaneously because an abrupt startup would expose all turbines simultaneously to the undisturbed background flow, unlike a developed wind farm in which downstream turbines experience the wakes of upstream turbines.

In contrast, Experiments 1 and 2 use precursor simulations that provide well-balanced initial wind fields, and the analysis targets the quasi-steady state long after the formal run starts (10 h and 6 h, respectively). Because the initial wind-farm transient is negligible in these cases, the ramp-up is omitted for simplicity.

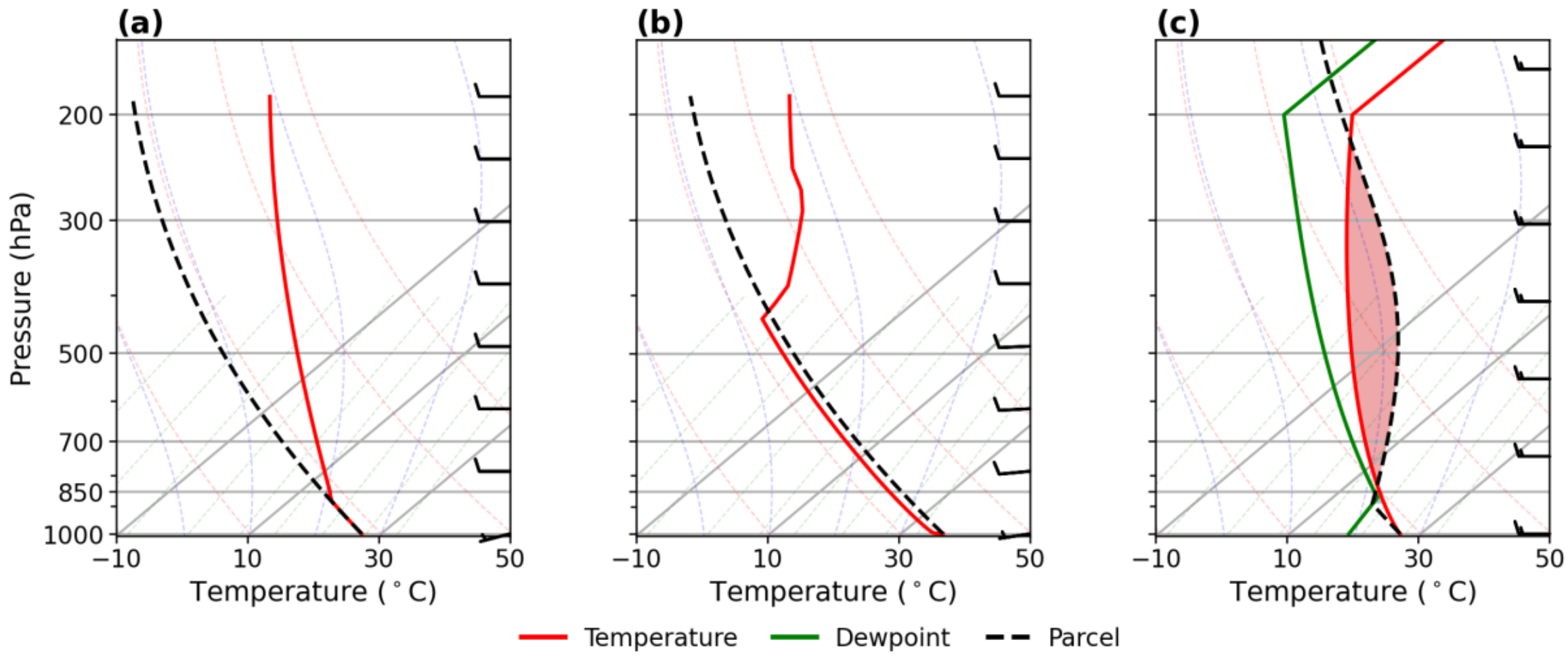

**Figure 2** The input soundings of formal simulations in Experiment 1 (a), 2 (b) and 3 (c)

# 3. Scheme Validation and Atmospheric Responses

## 3.1 WRF–CM1 Intercomparison (Experiment 1)

Direct comparison between CM1 and WRF simulations shows excellent agreement in the spatial distribution and magnitude of wind farm-induced perturbations. Both models reproduce the canonical signatures of a momentum sink and a TKE source that characterize wind farm wakes, as shown in Figure 3 and Figure 4. The downstream extent, spreading angle, and magnitude of the wind speed deficit and TKE enhancement are very similar in both the horizontal and vertical planes.

The maximum reduction in wind speed occurs near the downstream edge of the farm as the cumulative drag of the turbines takes effect. Downstream of the farm, the velocity deficit persists over a long recovery distance extending tens of kilometers downwind.

The maximum TKE enhancement occurs near the rotor-top level at the downstream end of the wind farm, which is attributed to direct turbine-generated TKE and shear production induced by the momentum deficit (Fitch et al., 2012).

The horizontal wind speed deficit and TKE enhancement are not confined solely to the rotor swept area but extend vertically throughout the depth of the boundary layer. Comparing the two, the wind speed deficit extends farther horizontally, whereas the TKE enhancement shows stronger vertical transport. Compared with the results of Fitch et al. (2012), the wake recovers faster and more TKE is advected downstream. We attribute this partly to differences in model setup and turbine specifications, and partly to the correction of the scheme from Archer et al. (2020).

Both models show a weak increase in wind speed in the near-surface layer within the wind farm. A reduction in TKE is observed in the near-surface layer downstream of the wind farm in both models, consistent with the reduced wind shear in the wake and the resulting reduction in shear production of TKE (Fitch et al., 2012).

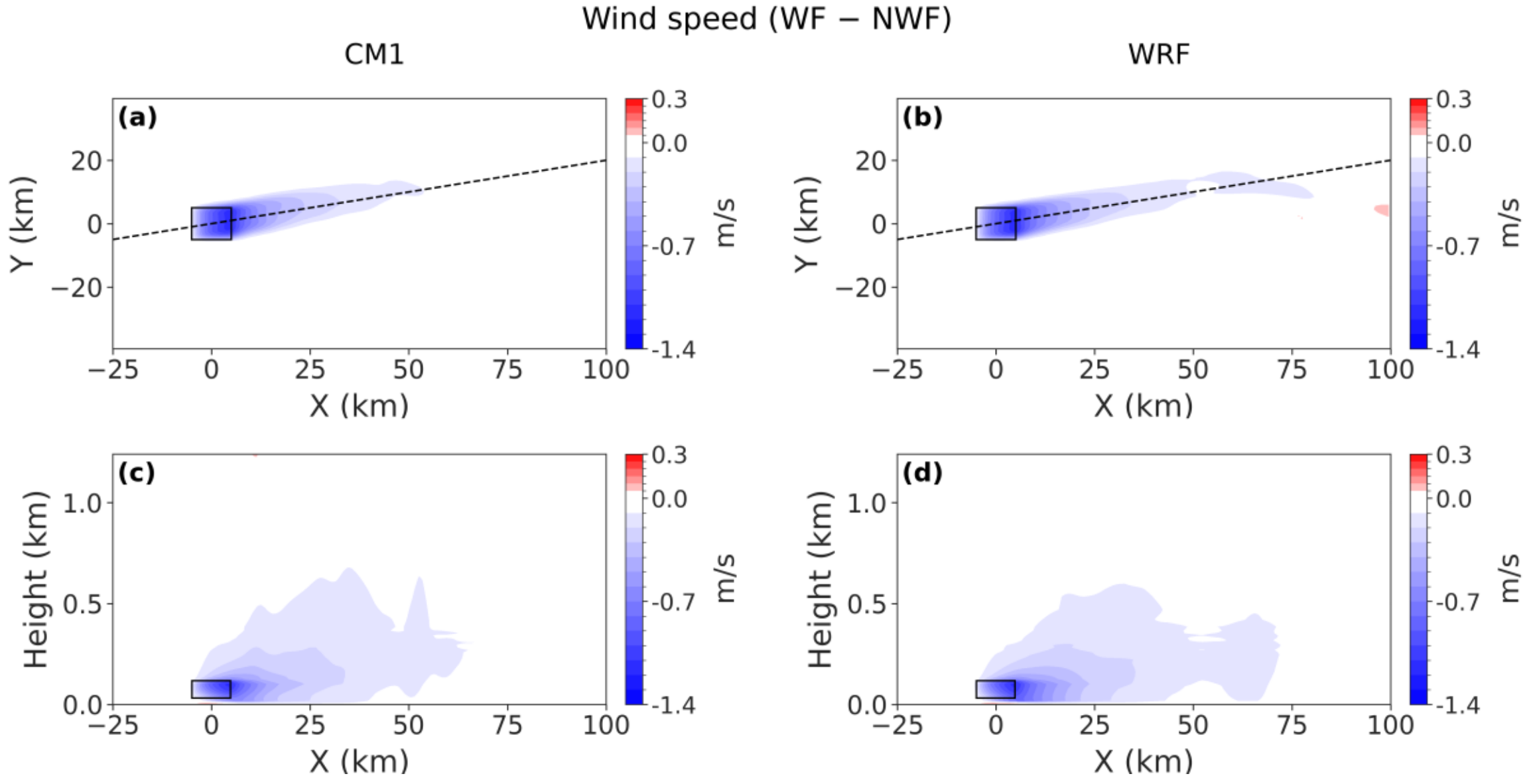

**Figure 3** The horizontal wind-speed difference (WF − NWF) at the final output time. Panels (a, c) show CM1 and (b, d) show WRF. Panels (a, b) are horizontal cross sections at 100 m, and (c, d) are vertical cross sections along the dashed line. The box marks the wind farm, and the origin is the farm center.

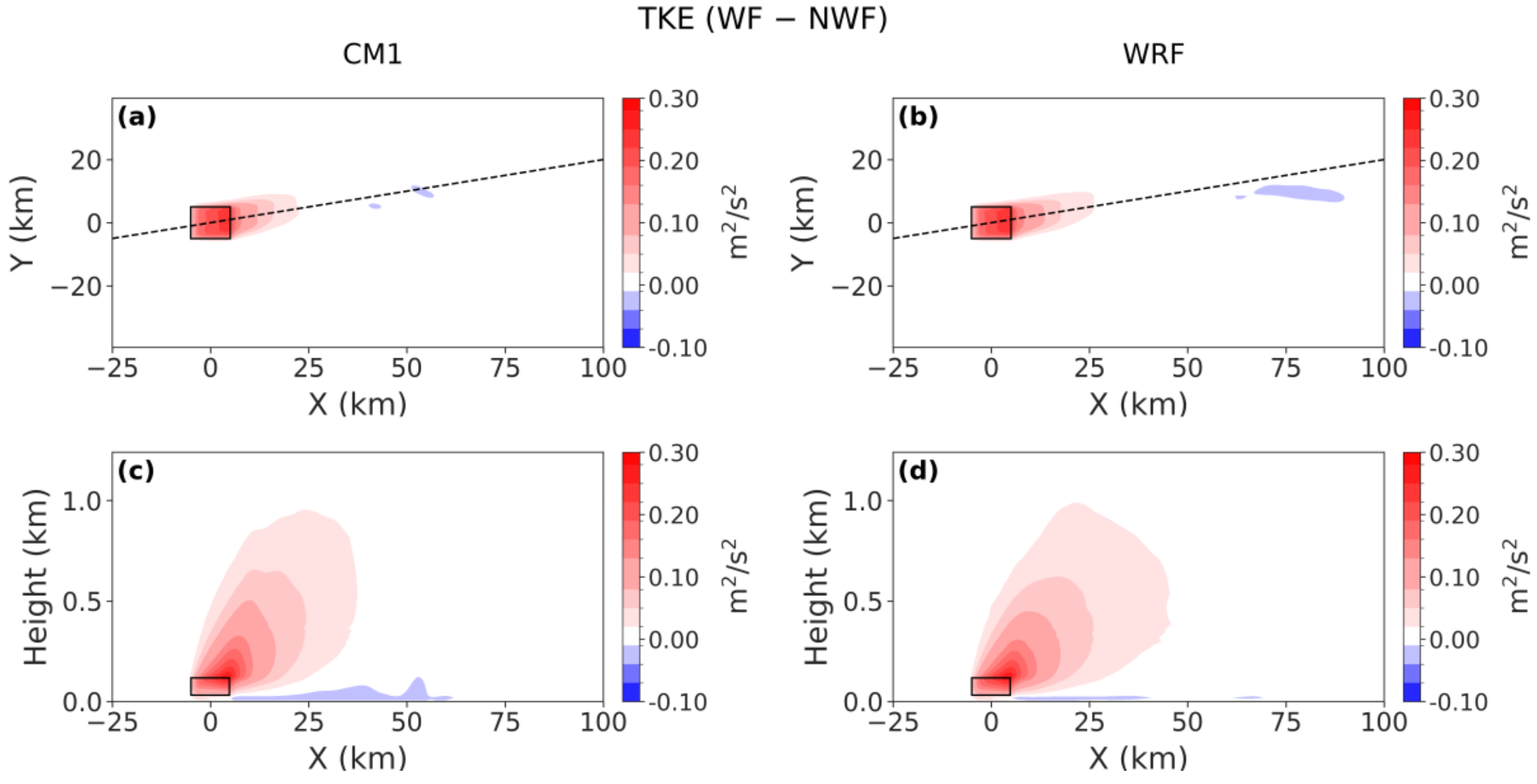

**Figure 4** The turbulent kinetic energy difference (WF − NWF) at the final output time, with the same layout as Figure 3.

## 3.2 Diurnal Cycle of Wind Farm Impacts (Experiment 2)

### 3.2.1 Wind, Turbulence, and Boundary-Layer Structure

As shown in Figure 5a and 5b, the wind-farm impacts show a pronounced day–night contrast. Under stable nighttime conditions, the wind speed reduction is confined almost entirely to the rotor layer. The wind farm becomes the dominant source of boundary-layer TKE, with TKE enhancements likewise concentrated within the rotor layer. Under unstable daytime conditions, the wind speed reduction is weaker but extends much higher into the boundary layer because of enhanced vertical mixing. TKE enhancements similarly extend to higher levels.

According to Figure 5a, the 2 MW wind-farm configuration weakens the nocturnal jet-like wind-speed maximum and shifts its core to a higher altitude. Under the 22 MW configuration, this structure disappears. Although the wind-speed maximum does not always strictly satisfy conventional criteria for a boundary-layer low-level jet (BLLJ), its evolution provides a useful measure of wind-farm-induced changes in the nocturnal wind profile. The wind farm also modifies the nocturnal Ekman spiral, extending the wind-direction adjustment to greater heights, with stronger turning aloft and weaker turning near the surface (Figure 6c). Relative to the no-wind-farm case, the 10-m wind speed is enhanced at night but reduced during the day (Figure 7a).

Overall, these results demonstrate a strong dependence of the boundary-layer response to wind-farm forcing on atmospheric stability.

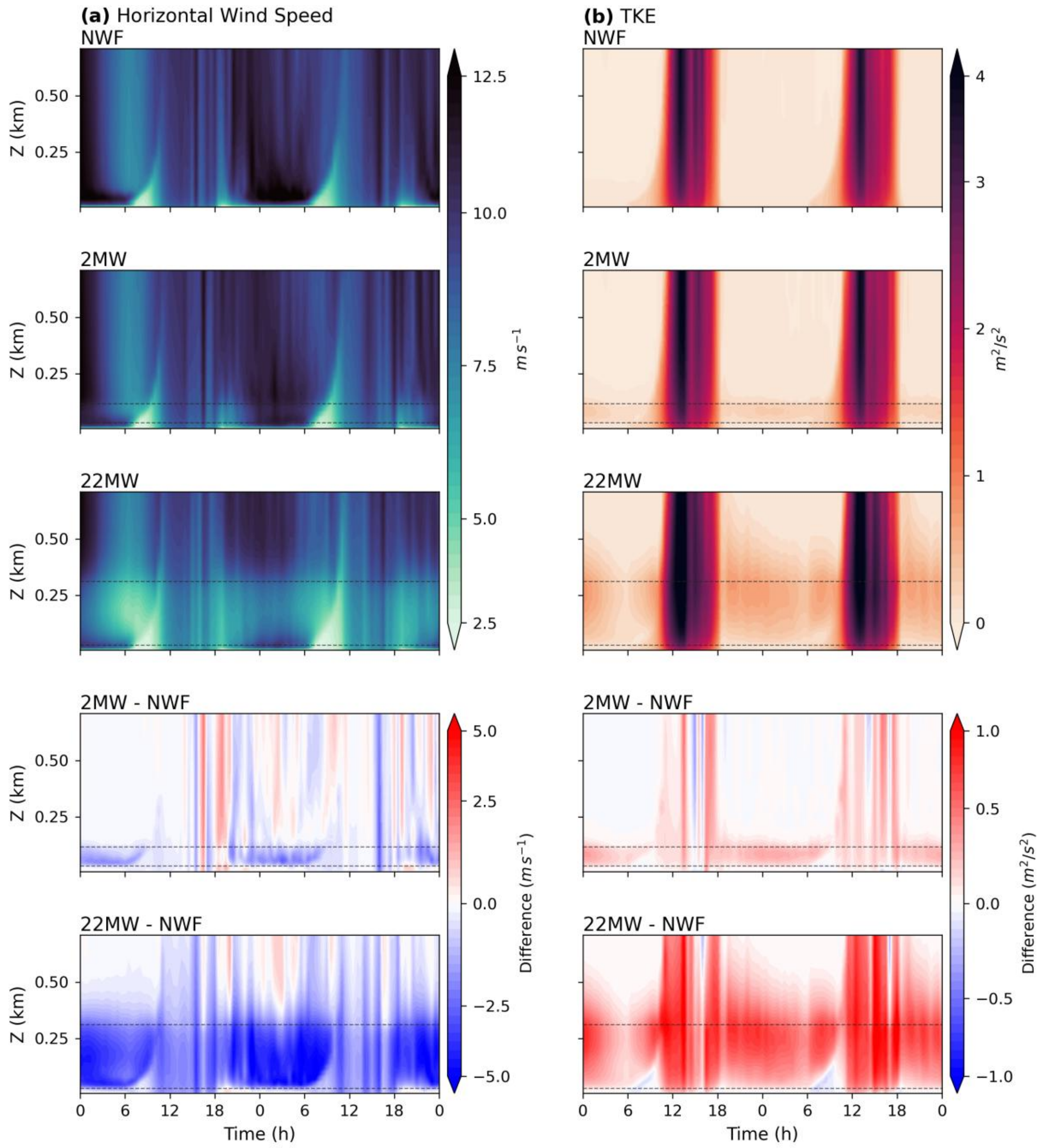

**Figure 5** Diurnal evolution of the wind-farm cases. (a) Horizontal wind speed and (b) turbulent kinetic energy as height–time sections for NWF, the 2 MW farm, and the 22 MW farm, and for the two farm-minus-NWF differences. Dashed lines mark the top and bottom of the turbine rotor.

### 3.2.2 Thermodynamic Processes

According to Figure 6a and Figure 7c, at night, wind farms induce near-surface warming and rotor-layer cooling, with temperature perturbations of approximately 0.5–1 K for the 2 MW configuration and 1–1.5 K for the 22 MW configuration. These changes weaken the inversion near hub height while strengthening the stratification above the rotor disk. During the morning transition, the wind-farm region becomes cooler than the surroundings at 2 m. Once the boundary layer transitions to an unstable regime, temperature perturbations induced by the wind farm become small. With the wind farm, the surface sensible heat flux increases only during the morning transition and decreases during the rest of the day (Figure 7e). This behavior is mainly associated with warmer near-surface air at night, cooler near-surface air during the morning transition, and weaker daytime winds. The contributions of turbulent mixing and vertical advection to local temperature changes are examined in Appendix C.

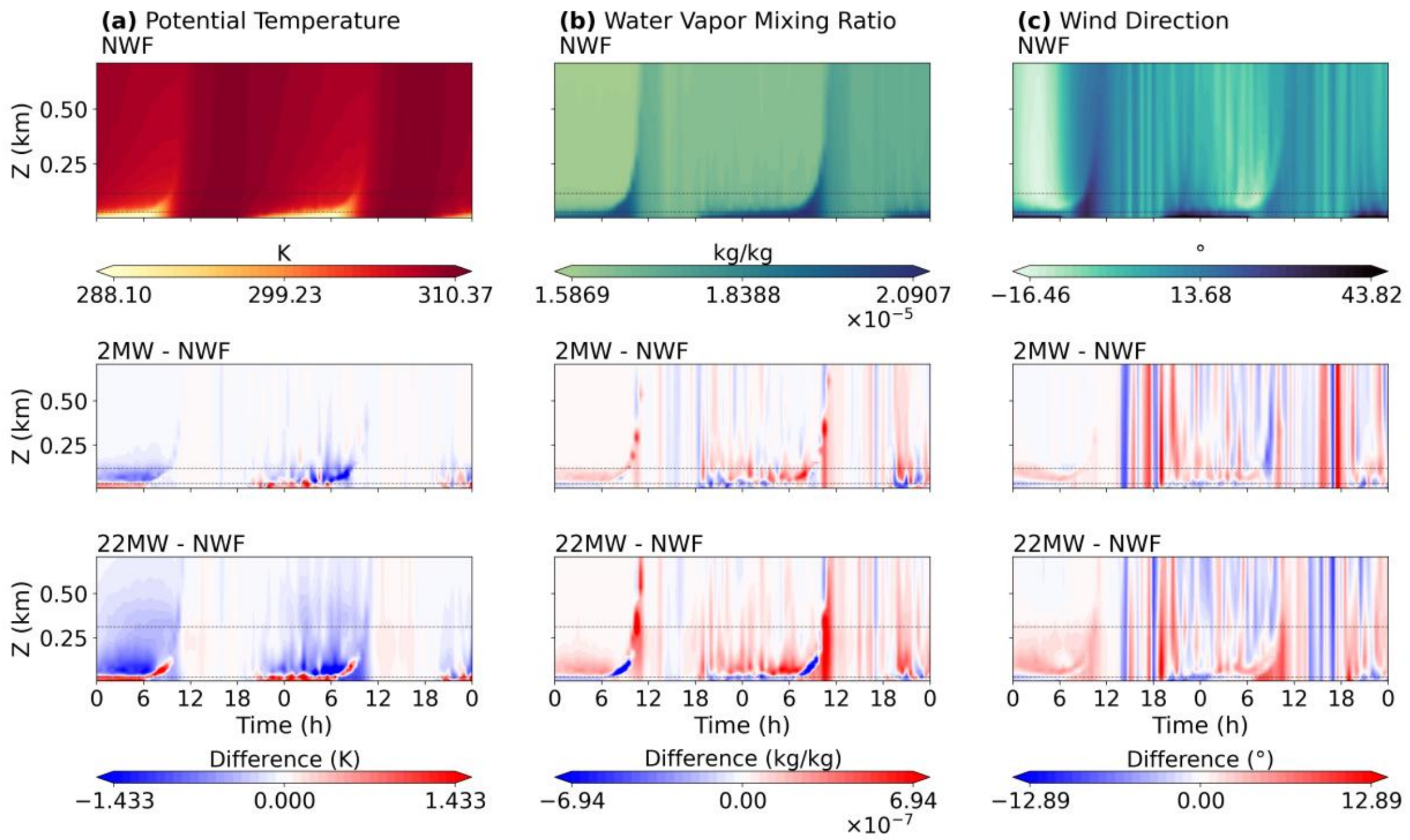

**Figure 6** Height–time sections of (a) potential temperature, (b) water vapor mixing ratio, and (c) wind direction, averaged over the wind farm region. The top row is the no-wind-farm control (NWF); the lower rows are the 2 MW and 22 MW farms minus NWF. Dashed lines mark the top and bottom of the turbine rotor.

As shown in Figure 6b and Figure 7d, under the experimental setup considered here, where atmospheric moisture is supplied primarily by local surface evaporation, wind-farm-induced mixing dries the near-surface layer and moistens the rotor layer at night. During the morning transition, the near-surface layer becomes anomalously moist. During the day, surface water-vapor flux decreases with the reduced near-surface wind speed, although this response may depend on the land-surface scheme and experimental setup (Figure 7f). Meanwhile, enhanced vertical mixing transports moisture away from the surface, with both effects contributing to a drier near-surface environment.

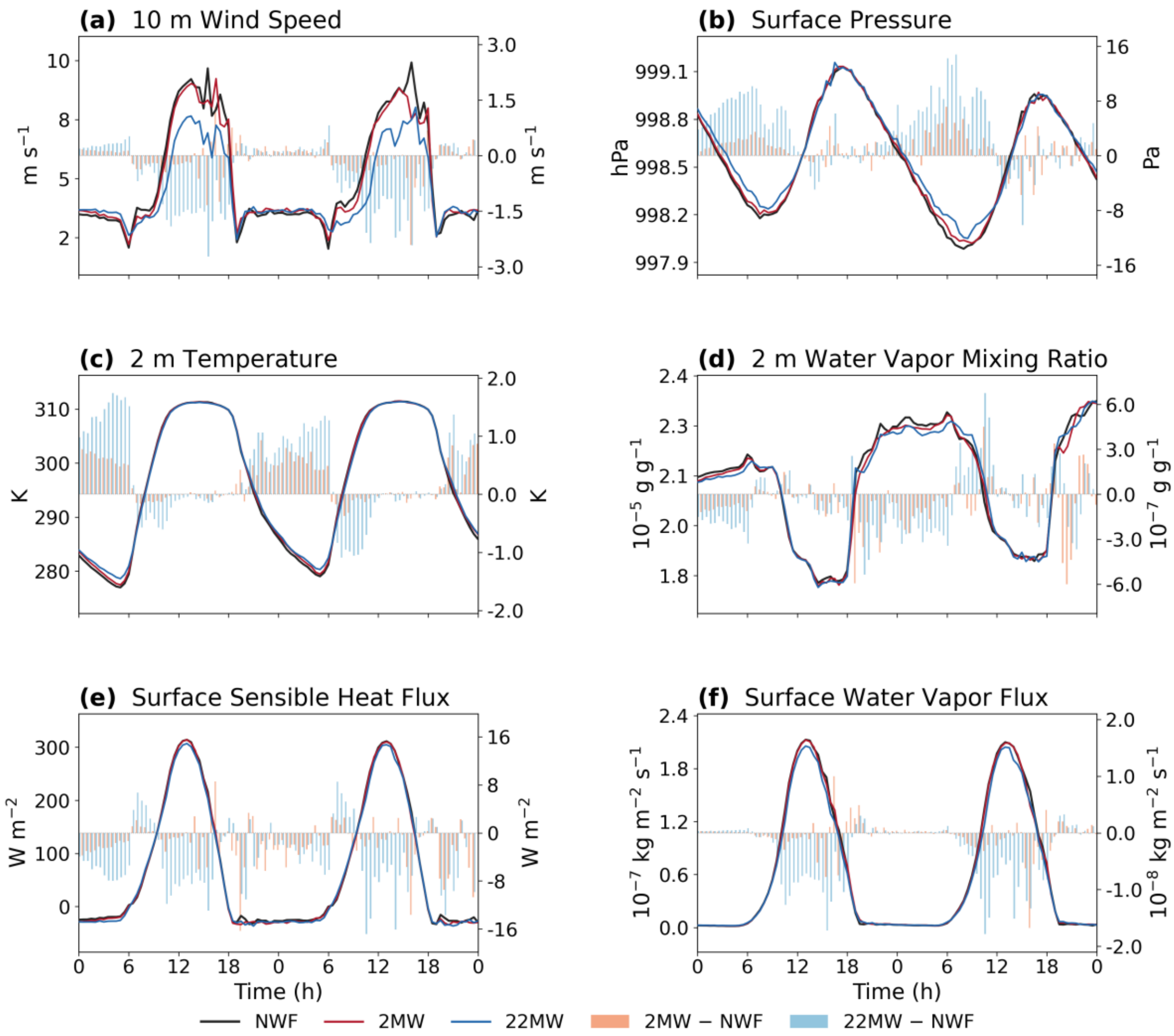

**Figure 7** Diurnal cycle of near-surface variables and surface fluxes averaged over the wind farm. Black, red, and blue curves are the no-wind-farm (NWF), 2 MW, and 22 MW simulations with the scale on the left axis. Bars are the differences from NWF (2 MW − NWF in orange; 22 MW − NWF in light blue), with the scale on the right axis. (a) 10 m wind speed. (b) Surface pressure. (c) 2 m temperature. (d) 2 m water vapor mixing ratio. (e) Surface sensible heat flux. (f) Surface water vapor flux.

The wind farm also produces small perturbations in surface pressure, generally on the order of a few pascals, with the largest increases occurring during the nighttime and transition periods (Figure 7b).

As expected, the dynamical and thermodynamic effects are generally stronger in the 22 MW case than in the 2 MW case.

## 3.3 Downstream Effects (Experiment 2)

In this section, we focus on the 22 MW case, where the signals are most pronounced. Related but generally weaker disturbances are also present in the 2 MW case.

### 3.3.1 Three-Dimensional Wake Structure and Vertical Motion

The pronounced wind veer within the nocturnal Ekman layer advects the wind-farm wake in different directions at different heights, producing a vertically skewed downstream wake (Figure 8a). Previous studies have shown that wind veer, together with the vertical mixing of veered momentum, can substantially deflect wind-farm wakes, particularly under stable conditions (Gadde & Stevens, 2019; van der Laan & Sørensen, 2017). Here, we examine how farm-scale wake deflection varies with height in mesoscale simulations. Our results show markedly different wake trajectories across turbine-relevant heights. The enhanced TKE is concentrated mainly along the upper edge of the velocity-deficit region (Figure 8b), consistent with enhanced shear production near the upper wake boundary.

The wind-farm wake also exhibits a pronounced asymmetry in vertical motion. At lower levels in Figure 8c, subsidence occurs predominantly on the right side of the wake viewed upstream, whereas ascent occurs on the left. Similar asymmetric vertical-motion pattern was reported by Fitch, Lundquist, et al. (2013) and attributed to wind-farm-induced modification of Ekman-layer flow. In addition, broader-scale vertical motions are present aloft in Figure 8c.

Under stable stratification, subsidence produces warming by advecting air with higher potential temperature downward, while ascent produces cooling. The temperature field below the rotor layer shows a warm–cold dipole and an additional cold anomaly along the outer flank of the wake (Figure 8d, 8e). Although turbulent vertical heat-flux divergence also contributes to the temperature tendency (Figure 8f), the potential temperature perturbations more closely follow the vertical advection tendency (Figure 8e).

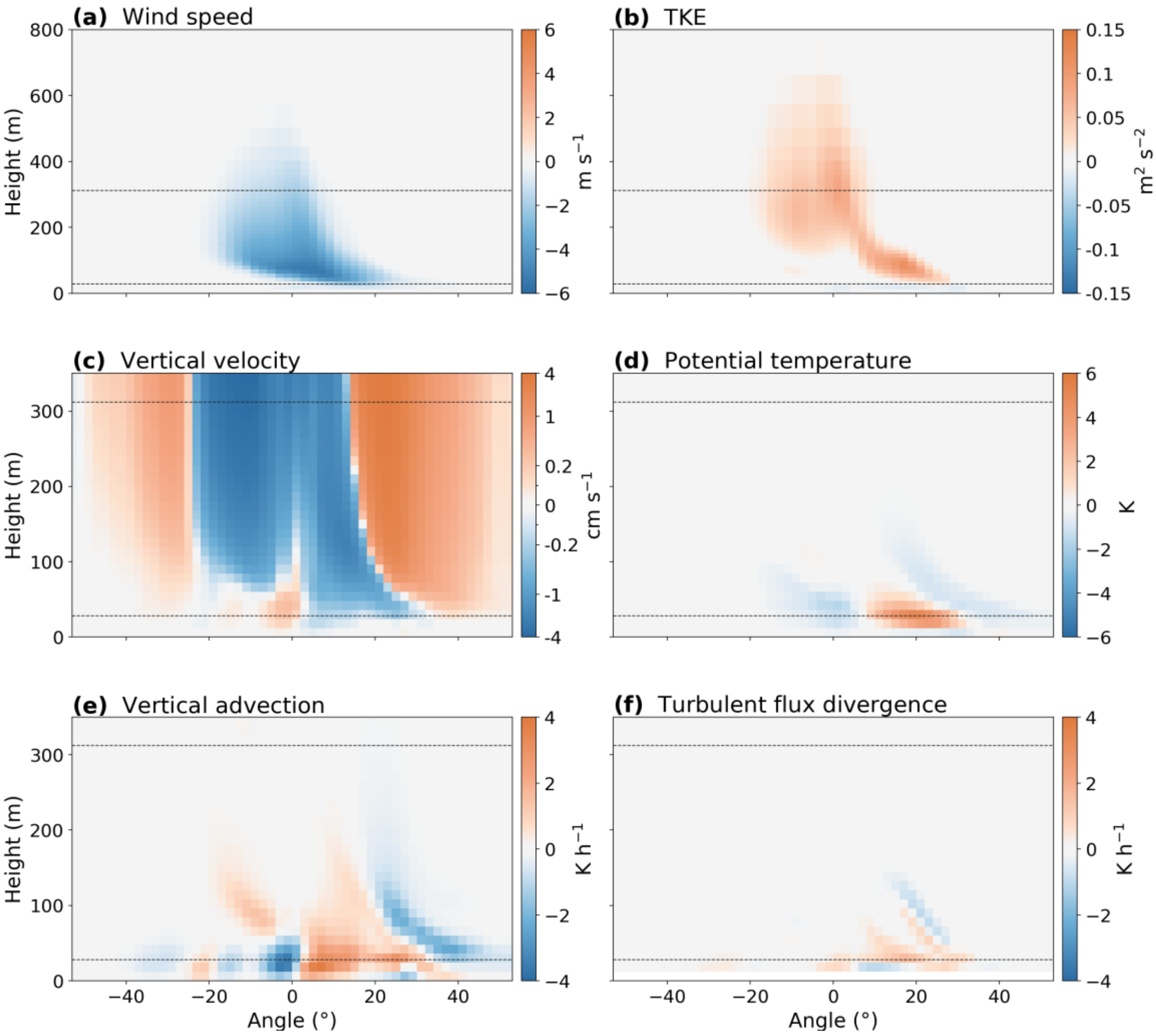

**Figure 8** Angle–height sections of the 22 MW minus no-wind-farm difference at 00:00 local time, averaged over 2° azimuth bins over the annulus 20–50 km from the farm. Azimuth is zero along the downstream axis. Dashed lines mark the lower and upper rotor tips (28 m and 312 m). (a) Horizontal wind speed. (b) Turbulent kinetic energy. (c) Vertical velocity. Its color scale is linear within $\pm\mathbf{0.2}\ \boldsymbol{cm\ s^{-1}}$ and logarithmic beyond that range. (d) Potential temperature. (e) Vertical advection of potential temperature, $-\boldsymbol{\Delta}(\boldsymbol{w}\frac{\partial\boldsymbol{\theta}}{\partial\boldsymbol{z}})$. (f) Turbulent heat flux divergence, $-\frac{\partial\boldsymbol{F}}{\partial\boldsymbol{z}}$, with $\boldsymbol{F} = -\boldsymbol{K_h}\frac{\partial\boldsymbol{\theta}}{\partial\boldsymbol{z}}$ and $\boldsymbol{K_h}$ the vertical eddy diffusivity for heat. Here $\boldsymbol{\theta}$ is potential temperature, $\boldsymbol{w}$ is vertical velocity, $\boldsymbol{z}$ is height, and $\Delta$ denotes the wind-farm minus no-farm difference. Positive values in (e) and (f) indicate warming ($\boldsymbol{K\ h^{-1}}$).

### 3.3.2 Near-Surface Response

During night and the morning transition, the wind farm may induce spatially extensive, organized perturbations in near-surface wind speed and temperature, affecting an area far larger than the wind farm itself. Although these perturbations are less apparent in the 2 MW case, they are substantially stronger in the 22 MW case and exhibit a distinct V-shaped structure (Figure 9a, 9b).

The background wind and vertical advective heating pattern in Figure 9b suggests that air parcels cross the heating and cooling regions more quickly along the lower wing than along the upper wing. This difference in crossing rate limits thermal accumulation along the lower wing but allows a larger temperature response along the upper wing, consistent with the temperature pattern (Figure 9c). The largest temperature anomalies can occur well downstream of the wind farm rather than within the farm itself. Because these anomalies are spatially localized and evolve with the background flow (Figure 9d), spatial or temporal averaging may underestimate their magnitude. Gravity waves and blockage effects may contribute to these organized perturbations.

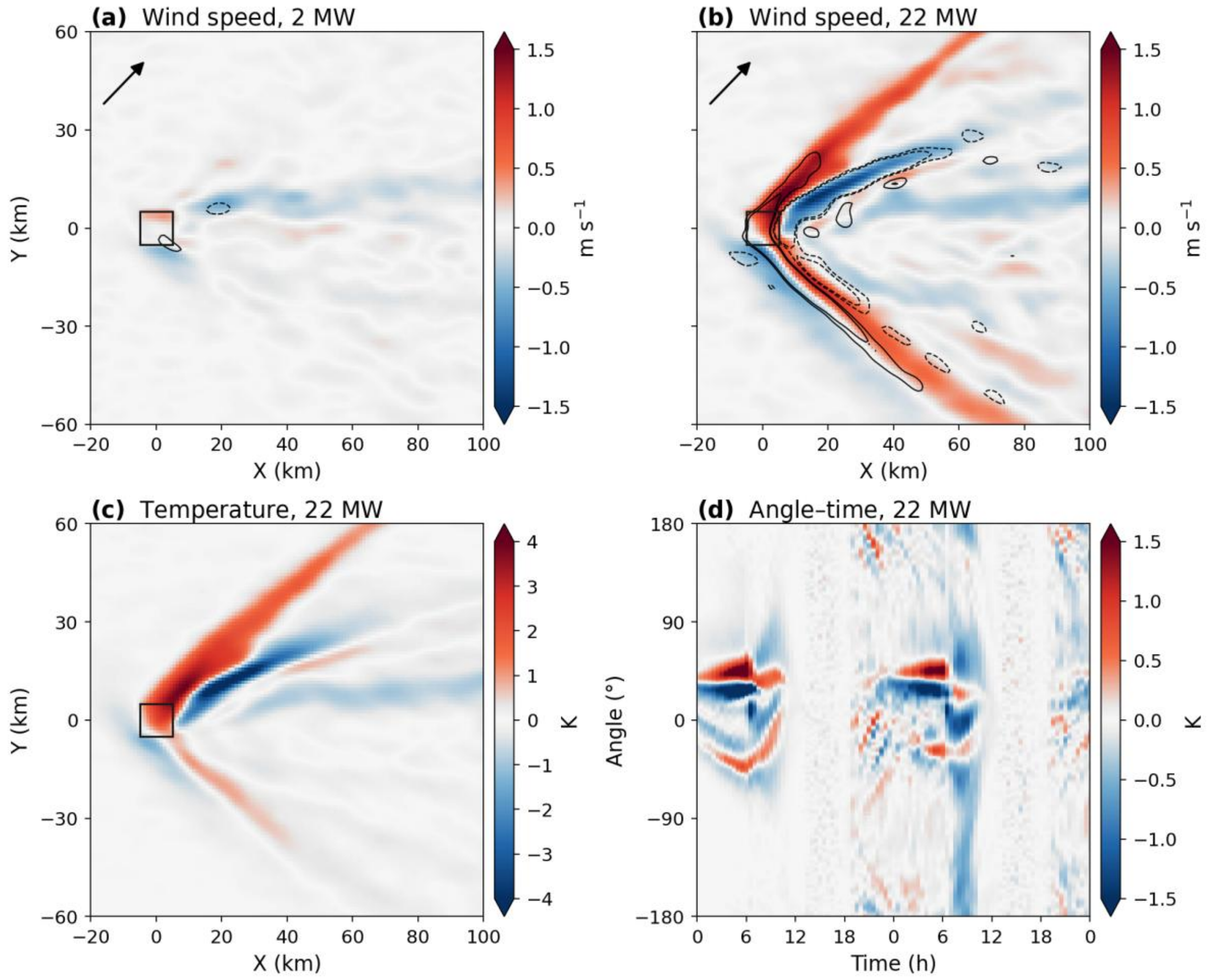

Figure 9 Near-surface wind, temperature, and the phase between wind and heating on the two downstream wings. (a) is 2 MW minus the no-wind-farm case; (b), (c), and (d) are 22 MW minus the no-wind-farm case. (a) (b) 10 m wind speed at 06:00 local time. The arrow is the 10 m background wind of the no-wind-farm case. Contours are the 6 m vertical advective heating ($K\ h^{-1}$), solid where positive and dashed where negative, at $\pm1$ and $\pm1.5\ K\ h^{-1}$. (c) 2 m temperature at the same time. The farm is marked by the box. (d) 2 m temperature, area-weighted from 20 to 50 km and over a 5° azimuthal sector, against azimuth and time from local midnight.

# 4. Impacts on Convection Initiation (Experiment 3)

## 4.1 Characteristics of the Convection Initiation

Under the favorable environmental conditions prescribed in Experiment 3, convection initiation (CI) exhibits a clear sensitivity to turbine rated power. Here we define CI as the first occurrence of composite radar reflectivity ≥35 dBZ (Mecikalski & Bedka, 2006). As shown in Figure 10, for the 15 MW and 22 MW configurations, convection initiates approximately 40–50 km downstream of the wind farm, well before the spontaneous initiation in the no-wind-farm (NWF) simulation, and subsequently develops into stronger, more organized convection. The 5 MW and 10 MW configurations also trigger downstream convection in regions where no corresponding initiation occurs in the NWF simulation, but the resulting convection is weaker and develops later than in the 15 MW and 22 MW cases. In contrast, the 2 MW configuration shows no clear systematic influence on CI. At later times, convection also develops about 15 km upstream of the wind farm in both the wind-farm and NWF simulations, suggesting that this upstream initiation is largely stochastic. We therefore focus on the wind-farm-induced downstream CI and use the 15 MW configuration as a representative case for the following analysis.

The cloud field shows a progressive evolution in the presence of the wind farm (Figure 11). Enhanced cloud development first appears over the wind-farm region, followed by stronger cumulus growth farther downwind. These clouds are then advected by the background flow and continue to intensify, culminating in CI approximately 40–50 km downstream of the wind farm.

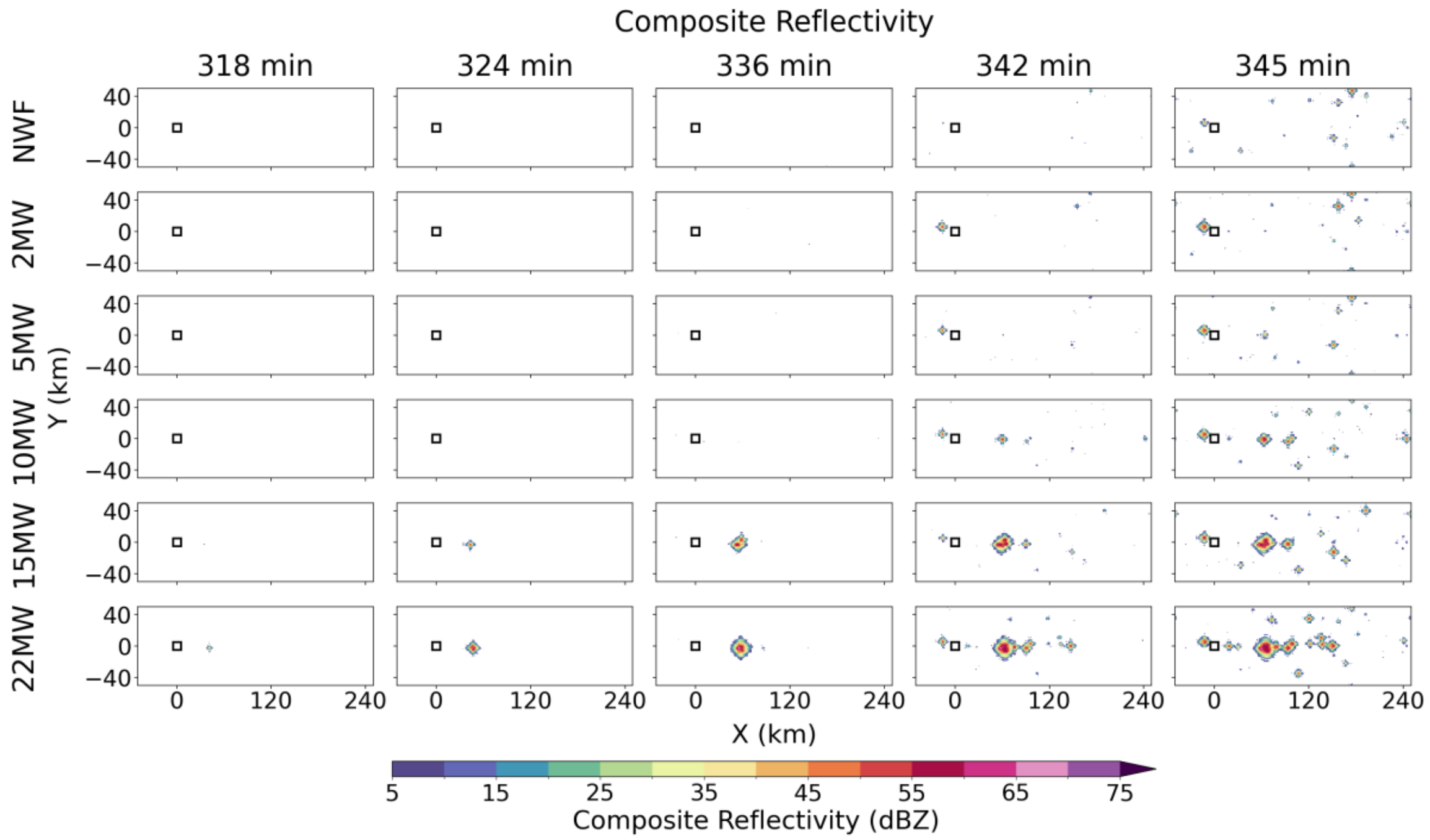

**Figure 10** Composite reflectivity for the NWF and 2, 5, 10, 15, and 22 MW cases from 318 to 345 min. The black box indicates the nominal wind-farm location.

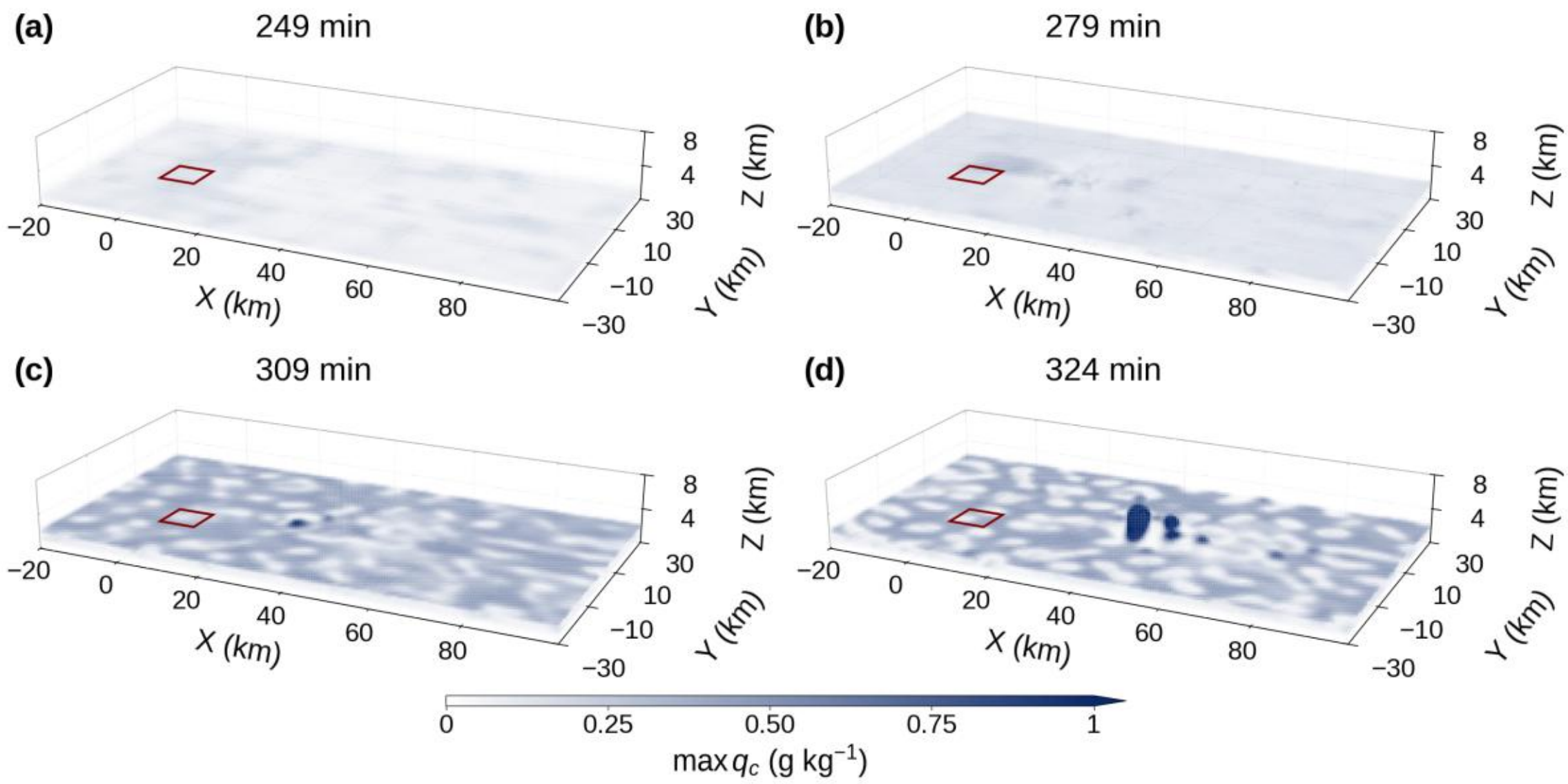

**Figure 11** Cloud-water mixing ratio at (a) 249, (b) 279, (c) 309, and (d) 324 min. Each colored pixel corresponds to a line through the three-dimensional domain, parallel to the viewing direction. The pixel color represents the highest cloud-water mixing ratio found anywhere along that line. A pixel is darkest blue if the mixing ratio reaches or exceeds 1 g kg$^{-1}$ anywhere along its line of sight. The maroon quadrilateral marks the outer edge of the 10 × 10 wind-turbine array. The vertical scale is exaggerated by a factor of two.

## 4.2 Physical Mechanisms of Convection Initiation

### 4.2.1 Gravity Wave Perturbation

The wind farm excites a vertically propagating gravity-wave response, characterized by alternating bands of ascent and descent, an approximately quadrature relationship between vertical velocity and potential-temperature perturbations, and phase lines that tilt upstream with height. The disturbance extends through the depth of the troposphere (Figure 12a). Convective perturbations first develop within the first ascending branch downstream of the wind farm (Figure 12b). They are subsequently advected by the background flow while progressively intensifying and deepening, eventually leading to CI (Figure 12c). This evolution links the CI to the ascending branch of the wind-farm-induced gravity wave.

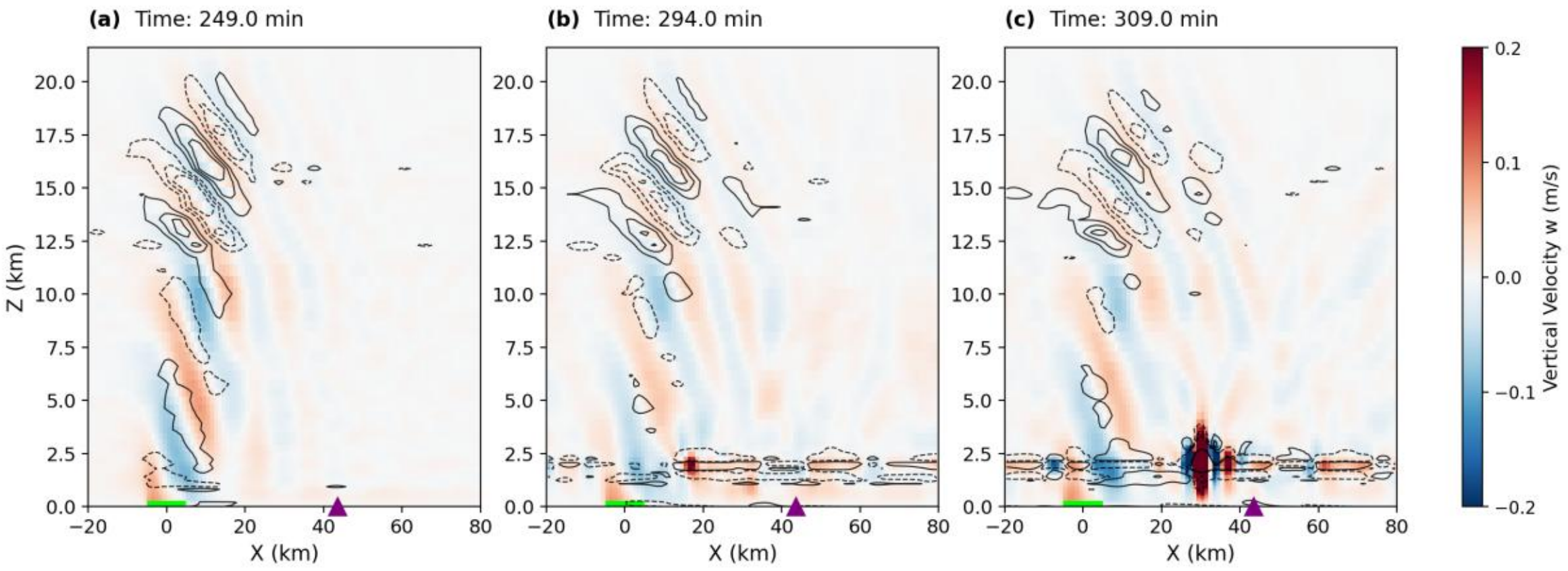

Figure 12 Vertical cross sections at y = −2.5 km showing the evolution of the turbine-induced gravity wave at 249, 294, and 309 min. Shading is vertical velocity and contours are potential temperature perturbation. The green bar marks the wind farm, and the purple triangle marks the downstream CI location at $x = 43.5\ km$.

### 4.2.2 Entrainment-Driven Boundary Layer Warming

Following turbine start-up, a persistent near-surface wind-speed deficit develops over the wind farm and extends downstream (Figure 13a). Within the wind-farm region, the reduced near-surface wind suppresses the surface water-vapor flux, reducing latent heat loss (Figure 13b). The surface energy balance consequently shifts toward sensible heating, leading to an enhanced sensible heat flux despite the weaker wind (Figure 13c). A positive potential-temperature anomaly develops over the wind farm and progressively extends downstream (Figure 13d). CAPE increases locally near the wind farm but decreases over much of the downstream region (Figure 13e), which is associated with the corresponding moisture changes (not shown). This spatial pattern suggests that downstream CI cannot be attributed simply to an increase in CAPE. Meanwhile, the background environment evolves toward higher CAPE and near-zero CIN, so CI ultimately occurs under negligible convective inhibition (Figure 13f).

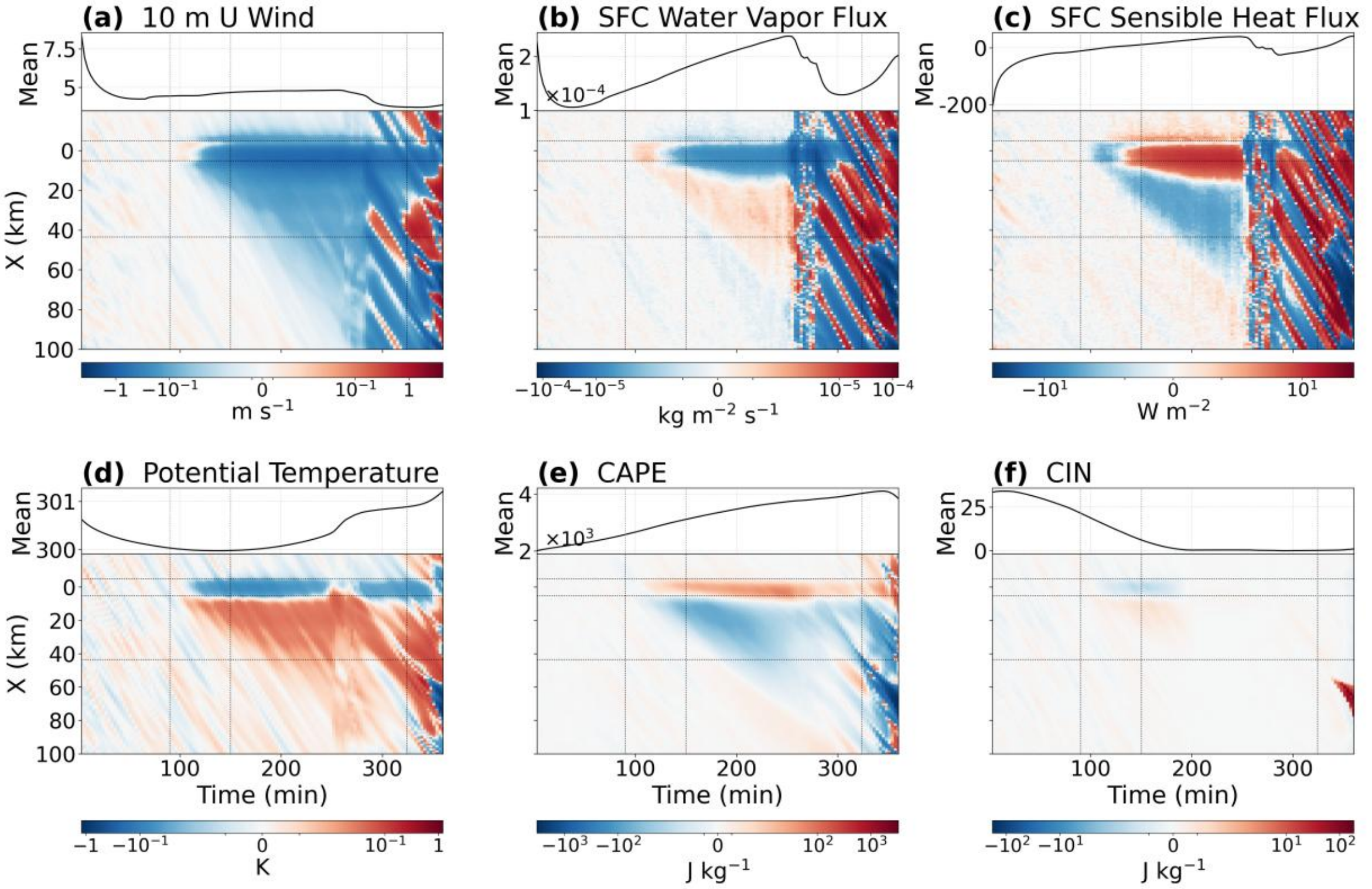

**Figure 13** Time evolution of near-surface and lower-tropospheric fields. In each panel, the upper curve shows the domain mean, while the lower field shows the streamwise departure from the mean, averaged across the wind farm (y = −5 to 5 km). Potential temperature is mass-weighted average over the lowest 1 km. Horizontal lines mark the upstream and downstream edges of the wind farm (x = −5 and 5 km) and the location of CI (x = 43.5 km). Vertical lines mark the start-up of the wind farm (90 min), the onset of its regular operation (150 min), and CI (324 min).

As shown in Figure 14a, the downstream warm anomaly develops into an elongated boundary-layer warm band aligned with the background flow, within which convection subsequently initiates. We therefore further examine other processes responsible for the formation of this warm band.

Figure 14b shows a stronger upward turbulent heat flux near the surface, consistent with the enhanced surface sensible heat flux in Figure 13c, and a much stronger downward turbulent heat flux in the upper boundary layer over and immediately downstream of the wind farm. The latter represents turbulent entrainment of warmer air from above into the boundary layer. The heat budget in Figure 14c quantifies the contributions of different processes and shows that this turbulent heat transport from boundary layer top, rather than surface heating or resolved advection, is the dominant source of the boundary-layer warming. The local turbulent mixing represented by the K-theory diffusivity accounts for most of this entrainment contribution. Notably, the resulting boundary-layer heating perturbation over the analysis region is about an order of magnitude larger than the total turbine power, indicating that the wind farm strongly reorganizes the ambient heat transport.

Once substantial cloud water develops above the boundary layer, the coupling among boundary-layer, surface, radiative, and microphysical processes complicates the potential temperature budget. By the period analyzed in Figure 14, however, the key thermodynamic and circulation conditions for subsequent CI had already been established.

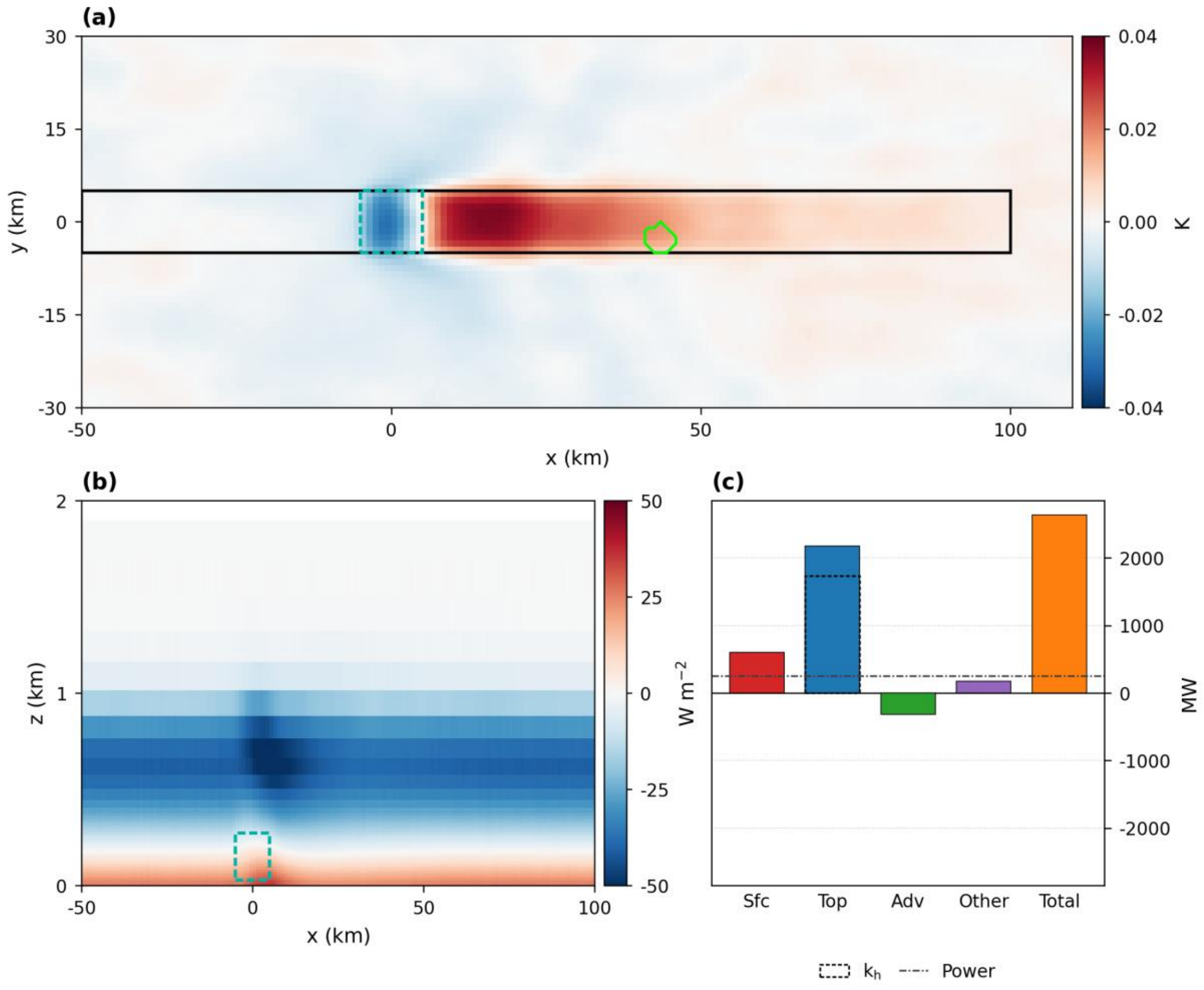

**Figure 14** Potential-temperature anomaly, mass-weighted over the surface–1 km layer and averaged over 210–270 min, relative to the domain-wide horizontal mean. The origin is the wind-farm center. The cyan dashed box marks the wind farm (x = −5 to 5 km, y = −5 to 5 km), the black rectangle marks the region used in (b) and (c) (x = −50 to 100 km, y = −5 to 5 km), and the lime contour outlines the CI region, defined by composite reflectivity ≥35 dBZ at 324 min. (b) Y-averaged turbulent heat flux H, diagnosed from the boundary-layer heating tendency as $\boldsymbol{H(z)} = \int_{z}^{z_{top}} \boldsymbol{\rho c}_{\mathrm{p}} \left(\frac{\partial \theta}{\partial t}\right)_{PBL} \boldsymbol{dz'}$, with $\boldsymbol{H(z_{top})} = \mathbf{0}$, and averaged over 150–240 min and y = -5 to 5 km. The cyan dashed box marks the wind-farm rotor layer (x = -5 to 5 km, z = 0.029 to 0.271 km). (c) Heat budget of the surface–1 km layer. Each bar is the departure from the domain-wide horizontal mean, integrated over 150–240 min and x = −50 to 100 km, y = −5 to 5 km. Sfc denotes $\boldsymbol{H(0)}$, while the blue bar is $-\boldsymbol{H(1km)}$ and excludes the surface flux. The dashed outline is recomputed from the K-theory flux at 1 km, $-\boldsymbol{\rho c_p K_h} \frac{\partial \theta}{\partial z}$, where $\boldsymbol{\rho}$ is air density, $\boldsymbol{c}_{\mathrm{p}}$ is the specific heat of air at constant

pressure, and $K_h$ is the eddy diffusivity for heat. Adv includes horizontal and vertical resolved advection, and Other includes the remaining processes. The dash-dotted line shows the total turbine power over the same area.

### 4.2.3 Downstream Secondary Circulations

Near the wind farm, ascent is concentrated over the upstream portion of the array, with subsidence immediately downstream. Farther downstream, the low-level vertical motion organizes into elongated bands approximately parallel to the background wind, with alternating ascent and descent accompanied by corresponding bands of horizontal convergence and divergence (Figure 15a–c). In cross section, these bands form adjacent counter-rotating circulations, giving the flow a roll-like structure, with low-level convergence beneath the ascending branches and divergence beneath the descending branches (Figure 15d–f). Two pronounced streamwise ascent bands emerge along the northern and southern flanks of the wind-farm wake and extend far downstream. The subsequent convection initiates along the southern branch at $x \approx 40 - 45$ km.

The transverse overturning constitutes a secondary circulation and can be understood as a response to lateral buoyancy gradients across the warm band produced by the wind farm. An idealized model of this response, together with its sensitivity to band geometry and layer depth and comparisons with the simulated circulation, is presented in Appendix D.

As the secondary circulation strengthens, the two ascending bands and their associated convergence intensify. Shortly before CI, localized strong updrafts develop along the preexisting bands, accompanied by locally much deeper cloud development. The strongest enhancement occurs along the southern band near the location of the subsequent convective cell, linking the evolving convergence and lifting structure to CI.

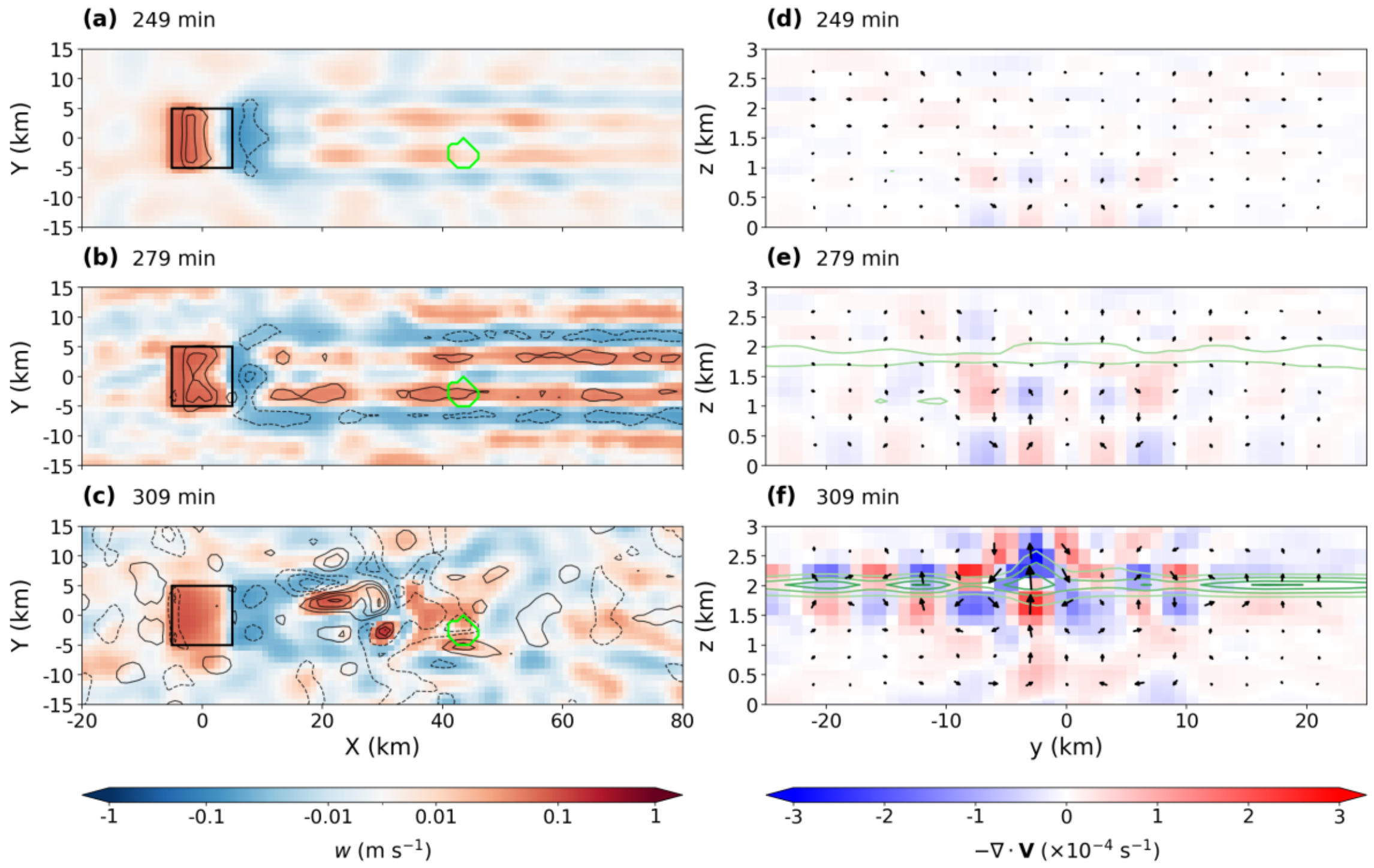

**Figure 15** Evolution of Downstream Ascent and Convergence. (a)–(c) Vertical velocity at 0.5 km (shading) and horizontal convergence at 10 m (black contours at $5 \times 10^{-5}$ $s^{-1}$ intervals; negative values indicate divergence). The rectangle outlines the 10 km × 10 km wind farm. The green contour shows the CI region. (d)–(f) Corresponding cross sections averaged over x = 23.5–43.5 km. Shading shows horizontal convergence ($10^{-4}$ $s^{-1}$; positive values indicate convergence). Green contours show cloud-water mixing ratio from 0.1 to 0.5 g $kg^{-1}$ at 0.1 g $kg^{-1}$ intervals, with darker green indicating larger values. Vectors show the cross-sectional wind components (v, w), plotted using the physical distance scaling of the axes so that arrow orientation represents the actual flow direction. Arrow length scales with the square root of wind speed.

### 4.2.4 Combined Effects on Convection Initiation

Along the downstream transect, the gravity-wave perturbation becomes collocated with the persistent ascending branch within the wind-farm wake, which provides sustained lifting through the boundary layer (Figure 16a–d). Although both the initial gravity-wave perturbation and the roll circulation have vertical velocities of only a few centimeters per second, the near-zero environmental CIN allows this weak lifting to support continued perturbation growth. As the disturbance is advected downstream, the associated upward motion strengthens and a localized moist anomaly develops near and above the boundary-layer top. Although the region downstream of the wind farm is relatively dry, low-level moisture-flux convergence associated with the secondary circulation compensates for this dryness and favors cloud formation and growth (Figure 16e–h). The perturbation therefore evolves from an initially weak wave-induced disturbance into a deeper and more localized convective updraft, eventually leading to CI.

Taken together, the downstream CI results from the combined evolution of thermodynamic and dynamical perturbations induced by the wind farm. The gravity wave provides the initial vertical perturbation, while turbulent entrainment modifies the downstream boundary layer and helps establish the secondary circulation. This circulation then sustains low level ascent and convergence, allowing the disturbance to grow and eventually initiate convection. The relative contributions of the gravity wave and thermodynamic effects remain to be further investigated.

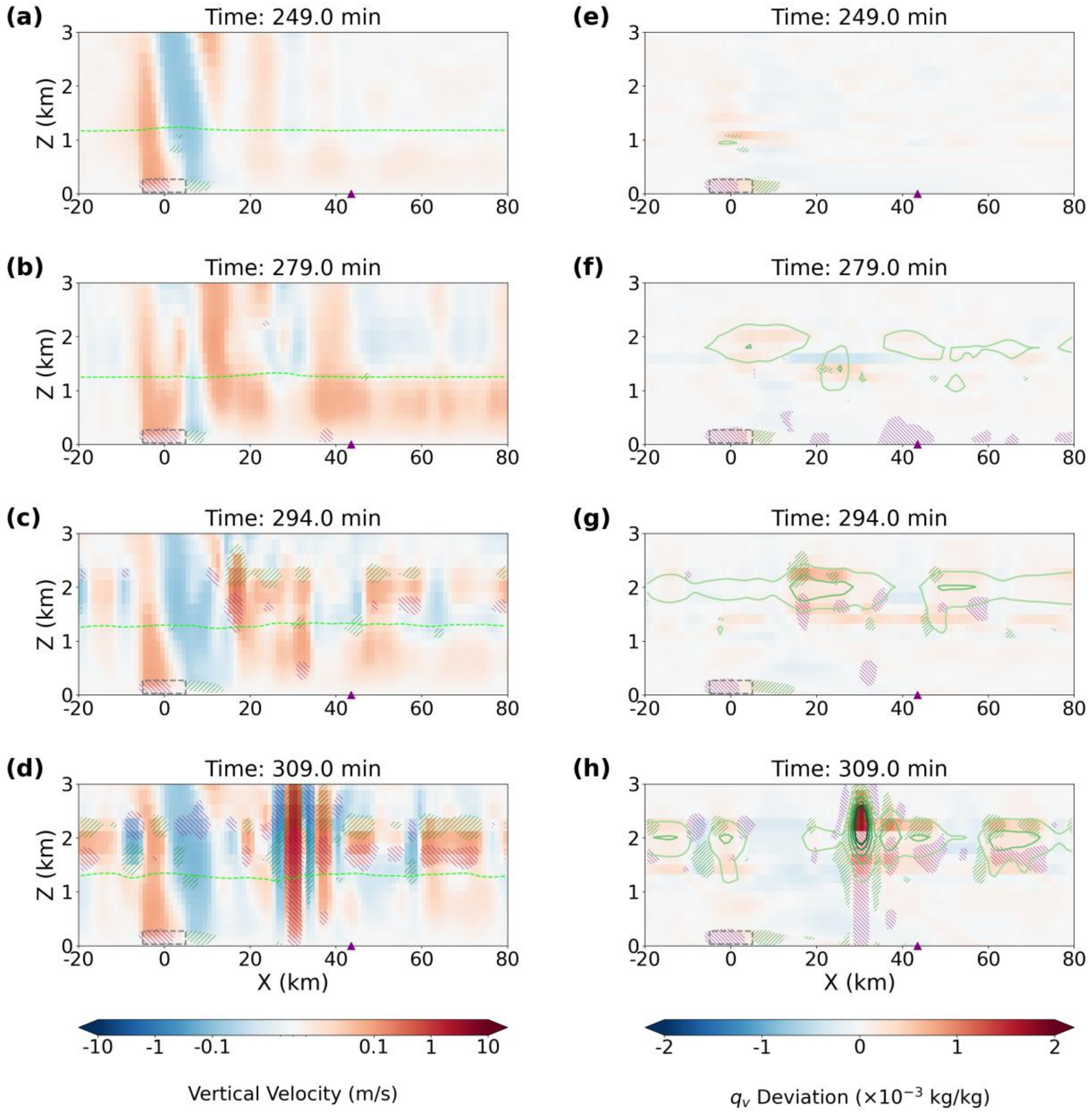

**Figure 16** **Vertical cross sections** at y = −2.5 km at 249, 279, 294, and 309 min. (a)–(d) Vertical velocity, with the dashed line indicating the planetary boundary layer height. Green and purple hatching denote horizontal divergence and convergence exceeding $1 \times 10^{-4}$ s$^{-1}$ in magnitude. (e)–(h) Water-vapor mixing-ratio deviation from the horizontal mean at each height. Green and purple hatching denote moisture-flux divergence and convergence exceeding $1.5 \times 10^{-6}$ kg m$^{-3}$ s$^{-1}$ in magnitude. Green contours show cloud-water mixing ratio from 0.1 to 0.5 g kg$^{-1}$ at 0.1 g kg$^{-1}$ intervals. The near-surface rectangle marks the wind farm, and the triangle marks the location of subsequent CI.

# 5. Discussion

## 5.1 Limitations of the Fitch Parameterization

A limitation of the present study is the use of the Fitch wind farm parameterization. Comparisons with LES show differences in turbulence induced by turbines and downstream wake recovery (García-Santiago et al., 2024; Peña et al., 2022). In situ evaluations further show that simulated wind speed and TKE can be sensitive to the choice of boundary layer scheme (Agarwal et al., 2026). More fundamentally, mesoscale wind farm parameterizations represent turbines as subgrid-scale effects rather than resolving individual turbine wakes, and generally require horizontal grid spacing of at least several rotor diameters (Fischereit et al., 2021). Further refinement toward sub-kilometer scales can enter the turbulence gray zone, where boundary layer motions become partially resolved and conventional boundary layer parameterizations face known limitations (Angevine et al., 2020). These limitations may affect the quantitative magnitude and fine scale structure of the simulated responses. Future work will therefore combine large eddy simulations with actuator disk turbine models and observational evaluation to assess the robustness of the mechanisms identified here and their quantitative effects.

## 5.2 Limitations of Idealized Simulation

The idealized setup also limits the direct applicability of our results to real world wind farm impacts. The simulated wind farm impacts, particularly their effects on CI, also remain to be validated against real-world observations. The simulations use horizontally uniform initial conditions and neglect terrain and surface heterogeneity, whereas realistic environments may contain stronger and more complex perturbations and interactions with other mesoscale systems, potentially altering both the boundary layer response and the timing and location of CI. In addition, the sensitivity experiments increase turbine power while holding turbine spacing and farm area fixed, which does not represent typical wind farm design. Future studies should therefore examine how power density, farm area, and layout affect the broader atmospheric response. Finally, only a single wind farm is considered here, while multiple nearby wind farms may produce cumulative or interacting effects that differ from those identified for an isolated farm.

## 5.3 Sensitivity to Initial Perturbations

Without the initial random temperature perturbations, wind farm related CI still occurs, but later, farther downstream, and in a more organized form (Appendix E). No CI occurs when both the wind farm and the perturbations are removed. Thus, the timing and location of CI are affected by the initial perturbations, while the occurrence of wind farm related CI remains relatively robust.

# 6. Summary

We implemented a Fitch-based wind farm parameterization in CM1, representing wind turbines as elevated momentum sinks and sources of turbulent kinetic energy (TKE) within the rotor-swept layer. The implementation includes optimizations for computational efficiency and memory use and supports OpenMP and MPI parallel execution. Three sets of idealized experiments were used to evaluate the implementation and examine wind farm effects on boundary-layer processes and downstream convection initiation.

Direct comparison with WRF shows close agreement in the magnitude and spatial structure of the wind-speed deficit and TKE enhancement. The diurnal experiments also reproduce established wind farm responses, including nocturnal near-surface warming and rotor-layer cooling, weakening and elevation of the nocturnal wind-speed maximum, and changes in Ekman turning. Wind-speed and TKE perturbations remain concentrated near the rotor layer at night but extend through a deeper layer during the day. Their consistency with previous studies provides additional support that the CM1 implementation captures the principal wind farm-atmosphere interactions represented by the Fitch scheme.

Under stable conditions, the downstream flow exhibits a vertically skewed wake and complex vertical-motion patterns. Near-surface wind-speed and temperature perturbations form a V-shaped pattern extending well beyond the farm. The temperature response is strongly asymmetric, and the largest temperature anomalies can occur downstream rather than within the farm. Because these anomalies are localized and evolve with the background flow, averaging over the farm or over long periods may understate their peak magnitude.

In the environment favorable for convection, the 15 MW and 22 MW configurations produce earlier and more organized convection approximately 40–50 km downstream than the no-farm simulation. The 5 MW and 10 MW configurations also promote downstream initiation, with a weaker and later response, whereas the 2 MW configuration shows no clear systematic influence.

Analysis of the 15 MW case indicates that convection initiation results from the combined evolution of a gravity-wave disturbance and a modified downstream boundary layer. Turbulent entrainment of warmer air across the boundary-layer top is the dominant source of the warming that forms an elongated downstream warm band, while the enhanced surface sensible heat flux makes a smaller contribution. Roll-like secondary circulations develop in the wake and provide persistent low-level convergence and ascent. The gravity wave induced by the wind farm supplies the initial vertical perturbation, which grows as it is advected downstream within an ascending roll branch. In an environment where convective inhibition is already near zero, sustained lifting and moisture-flux convergence promote cloud development and the eventual formation of a localized convective updraft.

# Appendix A: General Numerical Issues Encountered During Development

During development, we encountered several numerical issues that are not specific to the wind farm parameterization but affect all CM1/WRF users running certain physics configurations; they are documented here for the benefit of the broader community.

## A.1 RRTMG Radiation Buffer Layer Fix

The original CM1 implementation of the RRTMG longwave radiation scheme used a constant pressure decrement of 4 hPa to define the model-top radiation buffer layer, which represents the radiative effect of the atmosphere above the model domain. Because the number of buffer layers was determined from the initial-time model-top pressure, a fixed 4 hPa decrement could produce negative pressures in the buffer layer during long simulations or strong-convection events when the model-top pressure decreases, causing the model to crash. In WRF, the model-top pressure used by the RRTMG scheme is fixed by "p_top" and remains constant over time, while the pressures at all vertical levels are updated at each time step based on this value. This issue may cause runtime crashes in CM1 simulations using the RRTMG scheme.

Rather than adopting the complicated WRF approach, we applied a minor fix for experiments: the buffer-layer pressure decrement is now calculated dynamically at each radiation time step based on the actual model-top pressure, which ensures stability throughout the simulation.

## A.2 MYNN Level 3 Issue

The Fitch wind farm parameterization is only compatible with the Mellor–Yamada–Nakanishi–Niino (MYNN) boundary layer scheme. In CM1, this boundary layer scheme is available as two options: level 2.5 and level 3. The default MYNN configuration disables the "mix_qt" option, causing the total water counter-gradient term to be applied separately to both the water vapor and liquid water equations—effectively solving the counter-gradient correction twice in MYNN Level 3. We observed that this produces local anomalous negative or too large liquid water values in some vertical layers, which will force spurious condensation or evaporation. It can generate dry-hot or cold-wet perturbations and consequently trigger false convections in our simulations. However, the "mix_qt" option, which mixes total water instead of calculating them as individual species, is not yet mature enough to serve as a stable alternative (Olson et al., 2024). Previous studies have pointed out the potential systematic biases in the MYNN Level 3 scheme (Mantovani et al., 2026; Onwukwe et al., 2022; Prein et al., 2022), but whether the issues identified in this paper contribute to such errors requires further investigation. This issue is relevant to any user using MYNN Level 3 scheme in CM1, WRF and other models.

To ensure physically sound simulations, the MYNN Level 2.5 scheme is adopted for all experiments, which does not apply the counter-gradient term at all. This is consistent with previous wind farm simulations in WRF.

## A.3 Open Lateral Boundary Condition Artifacts

The radiation boundary condition (Durran & Klemp, 1983; Klemp & Wilhelmson, 1978; Orlanski, 1976) is applied at open lateral boundaries in atmospheric models such as CM1 and WRF. Boyer and Keeler (2022) documented a laminar inflow region in CM1 with little along-flow variability and large cross-flow variability. Earlier studies discussed limitations of lateral boundary treatments (Raymond & Kuo, 1984; Warner et al., 1997). The CM1 study (Boyer & Keeler, 2022) focuses on mitigating the laminar inflow region rather than deriving how initial potential-temperature variations at the boundary sustain the stripes. The normal-velocity nudging evaluated by Boyer and Keeler reduces the inward extent of the laminar region but does not directly modify the inflow potential-temperature field.

Appendix B examines this mechanism using a linear hydrostatic Boussinesq model under neutral stratification. The results indicate an incompatibility between random initial potential-temperature perturbations at the inflow boundary and the open radiative lateral boundary conditions considered here.

In high-precision ideal wind farm simulations, this artifact is particularly visible. To eliminate these artifacts, all experiments in the body of the text adopt fully periodic lateral boundary conditions with sufficiently large domains to ensure that the wake signal either decays before reaching the downstream boundary, or does not reach the downstream boundary at all during the simulation.

# Appendix B: Inflow Stripes in a Linear Hydrostatic Boussinesq Model

## B.1 Governing Equations and Pressure Decomposition

We consider a linear, hydrostatic, inviscid, adiabatic Boussinesq fluid under neutral stratification, with no Coriolis force and a uniform background wind $(U, V, 0)$, where $U > 0$. The vertical domain is $0 \leq z \leq H$. In the normal direction the domain is either $x \geq 0$ or $0 \leq x \leq L$; in the tangential direction it is periodic or bounded by two open sides. Here $x, y, z$ are the boundary-normal, tangential, and vertical coordinates, $H$ is the domain depth, and $t$ is time; $U$, $V$ are the background horizontal velocity components. Impermeable boundaries are imposed at the bottom and top, and the initial velocity perturbations vanish. Decompose the total fields as

$$\mathbf{u} = \mathbf{U}_0 + \mathbf{u}', \quad w = w', \quad p = p_0(z) + p', \quad \theta = \theta_0 + \theta'. \tag{B0}$$

Subscript 0 denotes the background state and primes denote perturbations. Here, $\mathbf{U}_0 = (U, V)$, $\theta_0$ and $\rho_0$ are constant. The background pressure $p_0(z)$ is hydrostatic, $\frac{\partial p_0}{\partial z} = -\rho_0 g$. Define buoyancy, kinematic pressure, and the linear material derivative as

$$b = \frac{g\theta'}{\theta_0}, P = \frac{p'}{\rho_0}, \qquad \frac{D}{Dt} = \frac{\partial}{\partial t} + U\frac{\partial}{\partial x} + V\frac{\partial}{\partial y} \tag{B1}$$

Let $u'$, $v'$, and $w'$ denote the perturbation velocities. In (B1), $g$ is gravitational acceleration, $\theta', p'$ are potential-temperature and pressure perturbations. The governing equations and vertical boundary conditions are

$$\begin{gathered} \frac{Db}{Dt} = 0,\ \frac{\partial P}{\partial z} = b, \\ \frac{Du'}{Dt} = -\frac{\partial P}{\partial x},\ \frac{Dv'}{Dt} = -\frac{\partial P}{\partial y}, \\ \frac{\partial u'}{\partial x} + \frac{\partial v'}{\partial y} + \frac{\partial w'}{\partial z} = 0, \\ w'(x,y,0,t) = w'(x,y,H,t) = 0 \end{gathered} \tag{B2}$$

An overbar denotes the vertical average over $0 \le z \le H$. A tilde denotes the zero-vertical-mean part of a velocity perturbation. Decompose the perturbation fields as $P = \bar{P} + \pi$, $u' = \overline{u'} + \tilde{u}$, and $v' = \overline{v'} + \tilde{v}$, with $\bar{\pi} = \bar{\tilde{u}} = \bar{\tilde{v}} = 0$. Here $\pi$ is the zero-vertical-mean baroclinic pressure and $\bar{P}$ is the barotropic pressure. Coordinate subscripts denote partial derivatives, and ${\nabla_h}^2$ is the horizontal Laplacian.

Column-integrated continuity and the vertically averaged momentum equations form the barotropic subsystem:

$$\begin{gathered} \overline{u'}_x + \overline{v'}_y = 0,\ \ {\nabla_h}^2 \bar{P} = 0, \\ \frac{D\overline{u'}}{Dt} = -\bar{P}_x,\ \frac{D\overline{v'}}{Dt} = -\bar{P}_y \end{gathered} \tag{B3}$$

The harmonic barotropic pressure $\bar{P}$ is determined by the horizontal boundary conditions. Hydrostatic balance and $\bar{\pi} = 0$ uniquely determine the buoyancy-induced baroclinic pressure:

$$\pi = -\int_z^H b(x,y,s,t)\, ds + \frac{1}{H}\int_0^H s b(x,y,s,t)\, ds \tag{B4}$$

The vertically uniform second term enforces $\bar{\pi} = 0$. Subtracting the vertically averaged momentum equations from (B2) gives $\frac{D\tilde{u}}{Dt} = -\pi_x$ and $\frac{D\tilde{v}}{Dt} = -\pi_y$.

## B.2 Lateral Boundaries and the Barotropic Response

We represent the open inflow treatment by radiation of the normal velocity and zero normal gradients of buoyancy and tangential velocity (Klemp & Wilhelmson, 1978). The effective normal propagation speed $C$ may be prescribed or diagnosed (Durran & Klemp, 1983); here it is vertically uniform and directed outward, $C \leq 0$. At $x = 0$, the boundary conditions are

$$\begin{gathered} b_x = 0, \\ u'_t + C u'_x = 0, \\ v'_x = 0. \end{gathered} \tag{B5}$$

This normal-wind update omits tangential advection and represents the CM1 treatment without the Coriolis force, diffusion, nudging, or other forcing. At a downstream boundary or an open tangential side, the normal component is likewise radiated outward and the tangential component has zero normal derivative. The radiation speed is smooth, bounded, and independent of height, but may vary with time, side, and tangential position.

The initial and boundary data are compatible, and the solutions are smooth. The barotropic velocity and pressure gradient are analytic in time. On the half-line, the perturbations and their time derivatives are uniformly bounded in the normal coordinate, and the barotropic fields have a spatially uniform time-analytic neighborhood.

Because the radiation speed is independent of height, the lateral velocity conditions pass to the vertical means. These means initially vanish. Suppose inductively that their time derivatives through order $m - 1$ vanish initially, starting with $m = 1$. Differentiating the mean momentum equations $m - 1$ times expresses the order-$m$ mean-velocity derivative as minus the gradient of the order-$m - 1$ barotropic pressure derivative; all advective terms vanish. Differentiating the normal radiation condition $m - 1$ times also makes the order-$m$ normal-velocity derivative vanish, since every product involving the radiation speed contains a lower-order velocity derivative. Hence this harmonic pressure derivative has homogeneous Neumann data on every open side.

For a finite normal interval, Green's identity makes the integral of the squared horizontal gradient of this pressure derivative zero: open-side fluxes vanish and periodic fluxes cancel. On the half-line, expand instead in tangential Fourier modes for periodic sides, or Neumann cosine modes for open sides. For each nonzero tangential wavenumber, boundedness removes the exponentially growing normal mode and the inflow Neumann condition removes the decaying mode. Boundedness also removes the linear part of the zero mode. The pressure gradient therefore vanishes in both geometries.

The order-$m$ mean-velocity derivatives consequently vanish, closing the induction. All time derivatives of both mean velocities and the barotropic pressure gradient vanish initially. Under the stated time-analyticity assumption, throughout the connected interval considered,

$$\overline{u'} = \overline{v'} = 0, \;\; \nabla_h \overline{P} = 0. \tag{B6}$$

The horizontal velocity perturbations therefore equal their baroclinic parts.

## B.3 Inflow Solution

Define $D_{\parallel} = \partial_t + V\partial_y$. At the inflow, the zero normal buoyancy gradient and the buoyancy equation give $D_{\parallel} b = 0$; the hydrostatic pressure closure gives $\pi_x = 0$. The tangential velocity condition then gives $D_{\parallel} v' = -\pi_y$. Defining $u'_b(y,z,t) = u'(0,y,z,t)$, the normal momentum and radiation conditions imply $\partial_t u'_b - \frac{CV}{U-C}\partial_y u'_b = 0$. Since $U - C > 0$ and the initial normal boundary velocity is zero, this homogeneous transport equation preserves that zero value wherever its backward boundary characteristic reaches the initial line without first entering from a tangential side.

In the boundary-influenced region $t > \frac{x}{U}$, a backward characteristic reaches the inflow at position $\eta$ and time $\tau$. Tracing the motion along the boundary back to $t = 0$ gives the initial coordinate $\xi$:

$$\eta = y - \frac{V}{U}x, \;\; \tau = t - \frac{x}{U}, \;\; \xi = \eta - V\tau = y - Vt \tag{B7}$$

Let $b_{ini}$ denote the initial buoyancy and $B(\xi, z) = b_{ini}(0, \xi, z)$ its boundary profile. With periodic tangential boundaries, the solution applies throughout the inflow region. With open sides, it applies where the backward characteristics, including their paths along the inflow boundary, reach the initial line without crossing a side. In this region, buoyancy conservation gives $b(x,y,z,t) = B(\xi,z)$. Equation (B4) determines the corresponding pressure $\pi(x,y,z,t) = \Pi(\xi,z)$. Integrating the baroclinic momentum equations along these characteristics gives

$$u' = 0, \;\; v' = -t\Pi_\xi(\xi, z). \tag{B8}$$

The tangential pressure gradient acts along both the boundary and interior portions of each characteristic, so acceleration accumulates over the total elapsed time. This response is driven by the buoyancy-induced pressure. Linear growth holds while perturbation advection remains negligible.

In contrast, the region where $t < \frac{x}{U}$ does not exhibit these organized motions. We define the inflow front as $x = Ut$, the position reached at time $t$ by air entering through the inflow boundary at $t = 0$ and advected at the background velocity.

## B.4 Vertical Velocity and Interpretation

The barotropic horizontal divergence vanishes by (B3), so continuity gives $w'_z = -\tilde{u}_x - \tilde{v}_y$. Substituting (B8) and integrating upward from $w'(x, y, 0, t) = 0$ gives

$$w'(x, y, z, t) = t\partial_\xi^2 \int_0^z \Pi(\xi, s)\, \mathrm{ds}. \tag{B9}$$

Equation (B8) gives zero normal velocity throughout the inflow subregion. The vertical velocity in (B9) consequently has no normal-coordinate dependence; its tangential variations form boundary-normal bands.

The zero vertical means ensure the top boundary condition:

$$\begin{gathered} \int_0^H \tilde{u}\, ds = \int_0^H \tilde{v}\, ds = \int_0^H \Pi\, ds = 0, \\ w'(x, y, H, t) = 0. \end{gathered} \tag{B10}$$

Successive inflow parcels carry the boundary's temperature contrasts into the domain. Because the boundary pattern and interior air move at the same tangential speed, these contrasts form boundary-normal bands even when the background wind is oblique. With tangential wind, the thermal and pressure bands translate along the boundary while the circulation strengthens. Their widths and spacing follow the initial boundary pattern. Hydrostatic pressure gradients drive cross-band flow, with vertically integrated convergence producing ascent and divergence producing descent. The vertical-motion extrema depend on pressure curvature and vertical structure.

## B.5 Comparison with a CM1 Simulation

Figure 17 compares the inflow solution with a shallow neutral CM1 integration. The domain is 50 km square and 300 m deep, with a uniform westerly background wind of 10 m s$^{-1}$, a potential temperature of 300 K, and no Coriolis force. The northern and southern boundaries are periodic; the eastern and western boundaries are open, with outward radiation of the normal velocity. A random potential-temperature perturbation of amplitude 0.25 K fills the domain, including the western boundary. The integration lasts 18 min.

West of the front, the simulated vertical velocity reproduces the stripe pattern of the inflow solution (Figure 17a, 17b). The profile at the sampled column agrees in shape, although its amplitude is slightly smaller (Figure 17c). After the front arrives, the tangential- and vertical-velocity standard deviations grow approximately linearly, as predicted by (B8) and (B9), while the normal-velocity perturbation approaches zero, as predicted (Figure 17d).

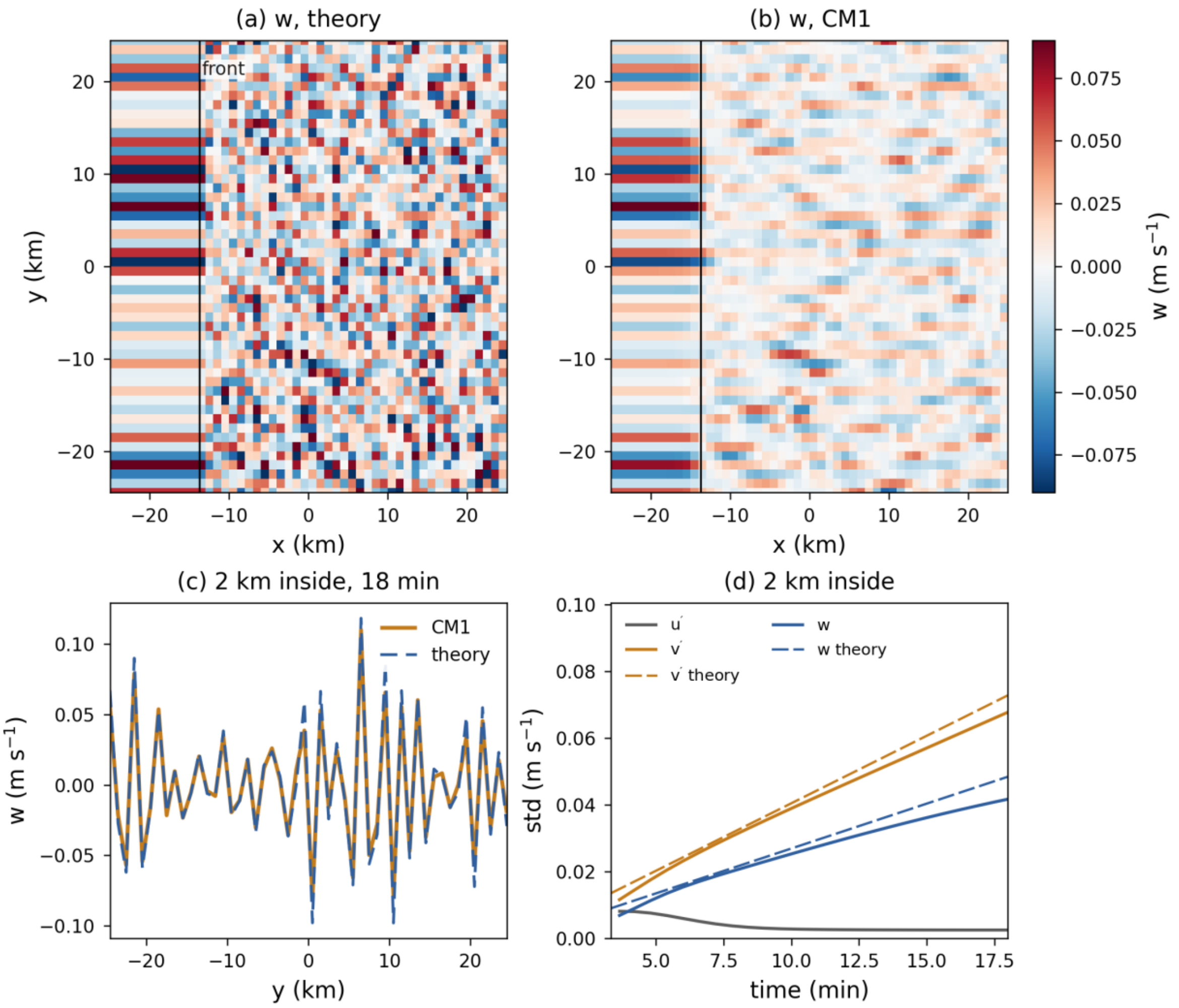

**Figure 17** Comparison of the inflow solution with a shallow neutral CM1 simulation at 150 m. The sampled column is 2 km east of the westernmost scalar point. At the imposed background speed, the inflow front reaches it after approximately 3.3 min and is 10.8 km east of the western boundary at 18 min. The theoretical profiles use the initial western-boundary buoyancy with its full vertical structure. Equation (B4) evaluates the baroclinic pressure over the full 300 m depth. The vertical velocity is calculated by integrating the tangential pressure curvature from the surface to that level over the model layers. Panels (a) and (b) show theoretical and simulated vertical velocity, respectively, at 18 min over the full horizontal domain; the black line marks the inflow front. Panel (c) shows the meridional profile at 18 min, 2

km inside the western boundary, and (d) shows the subsequent evolution of the meridional standard deviations at that column. Solid curves denote CM1 and dashed curves denote theory. The inflow solution applies west of the front.

## Appendix C: Local Heat Budget near the Wind Farm

The heat-budget analysis further clarifies the mechanisms underlying the temperature changes discussed in Section 3.3.2. Turbine-enhanced mixing produces a concentrated warming tendency near the surface, consistent with the conventional interpretation of nocturnal wind-farm warming (Fitch, Lundquist, et al., 2013). The relative role of organized advection remains less settled, with previous numerical and observational studies finding strongly varying contributions from advection and turbulent heat-flux convergence (Wu & Archer, 2021; Wu et al., 2024; Xia et al., 2019).

As shown in Figure 18, in our experiments, changes in surface heat flux are small, while turbulent mixing and resolved-scale advection both play important roles in the near-surface temperature response. Within the lowest 36 m over the wind farm, turbulent mixing provides the dominant heating contribution. Vertical advection, primarily associated with subsidence, also contributes to warming, whereas horizontal advection acts predominantly as a cooling term, offsetting the warming associated with turbulent mixing and vertical motion. Consequently, the net three-dimensional advective contribution is cooling and approximately balances turbulent heating in the time mean. Farther downstream, however, resolved-scale advection becomes the dominant contribution to the temperature tendency, consistent with previous studies (Xia et al., 2019).

Notably, despite substantially stronger near-surface warming in the 22 MW case than in the 2 MW case, neither turbulent mixing nor the net advective contribution over the wind farm increases appreciably. The most pronounced difference is instead found in the vertical advection contribution, which increases substantially in the 22 MW case.

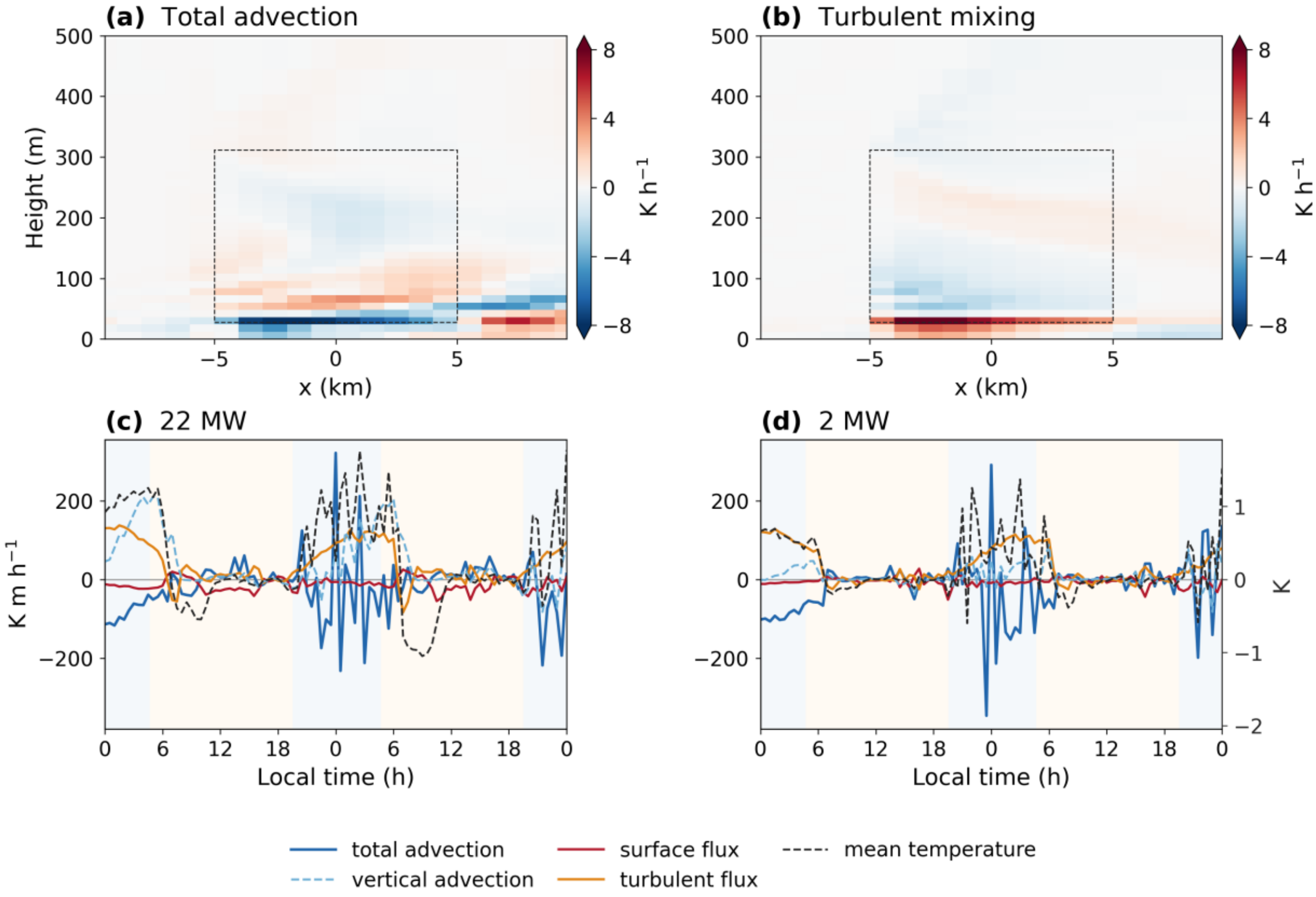

**Figure 18** Heat-budget perturbations relative to the no-farm simulation. (a, b) Total advection and turbulent mixing tendencies at 00:00 local time, averaged in the crosswind direction, from 5 km upstream to 5 km downstream of the wind farm. Total advection is the sum of the vertical and horizontal contributions. The dashed box denotes the farm extent and rotor layer (28–312 m). (c, d) Budget terms integrated over the lowest 36 m and averaged over the farm for the 22 MW and 2 MW cases. The solid blue line is the total advection, and the dashed blue line is its vertical component. The integrated turbulent term is decomposed into the surface heat flux and the turbulent heat flux at 36 m. The dashed black line shows the 0–36 m mean potential-temperature difference (right axis). Positive budget values indicate warming relative to the no-farm simulation, and positive temperature values indicate a warmer column. Cream and blue shading denote daytime and nighttime, respectively.

# Appendix D: Forced response of a finite-width warm band

We derive the steady, single-mode circulation forced by a finite-width warm band, examine its sensitivity to band geometry and layer depth, and compare its spatial structure and amplitude with the 15 MW simulation.

## D.1 Theoretical framework and forced response

Let WF and CTL denote the wind-farm and control states, respectively. Anomalies are

$$(\cdot)' \equiv (\cdot)_{\mathrm{WF}} - (\cdot)_{\mathrm{CTL}}. \tag{D1}$$

With $\theta$ the potential temperature and $q_v$ the water-vapor mixing ratio, virtual potential temperature and buoyancy are

$$\theta_v = \theta\,(1 + 0.61\,q_v), \qquad b = \frac{g}{\theta_0}\,\theta_v', \tag{D2}$$

where $\theta_0$ is the reference potential temperature and $g$ is gravitational acceleration. The coordinates $x$, $y$, and $z$ are directed along the band, across the band, and upward, respectively. The coordinate origin is the farm center. Here, $w'$ and $v'$ denote the vertical and cross-band velocity differences between the WF and CTL simulations, respectively.

Let H denote the roll depth. The reduced model sets $w' = 0$ at $z = 0$ and at $z = H$. The gravest structure is $\sin(mz)$ with $m = \frac{\pi}{H}$. The first-mode amplitudes of buoyancy, vertical velocity, and crosswind velocity, denoted by $B_1$, $W_1$, and $V_1$, respectively, are defined by

$$B_1 = \frac{2}{H}\int_0^H b\,\sin(mz)\,\mathrm{d}z, \tag{D3}$$

$$W_1 = \frac{2}{H}\int_0^H w'\,\sin(mz)\,\mathrm{d}z, \tag{D4}$$

$$V_1 = \frac{2}{H}\int_0^H v'\,\cos(mz)\,\mathrm{d}z. \tag{D5}$$

Let $B, W, V$ denote the buoyancy, vertical-velocity, and crosswind-velocity modal amplitudes. The subscript 1 in (D3)–(D5) labels diagnostic projections, whereas the unsubscripted amplitudes are used in the single-mode model. Retaining only the first vertical mode, $b = B\sin(mz)$, $w' = W\sin(mz)$, and $v' = V\cos(mz)$. Continuity, $\frac{\partial v'}{\partial y} + \frac{\partial w'}{\partial z} = 0$, then requires

$$\frac{\mathrm{d}V}{\mathrm{d}y} = -m\,W. \tag{D6}$$

Near the surface $\cos(mz) \approx 1$, so $v' \approx V$; at $z = H$, $\cos(mz) = -1$ and the cross-wind flow reverses.

In the steady, linear balance, variations along the roll are neglected and turbulent momentum loss is represented by drag of frequency $\gamma$ in both momentum equations,

$$\gamma v' = -\frac{\partial P}{\partial y}, \tag{D7}$$

$$\gamma w' = -\frac{\partial P}{\partial z} + b, \tag{D8}$$

where $P = \frac{p'}{\rho_0}$ is the perturbation kinematic pressure, with constant reference density $\rho_0$. Cross-differentiation removes $P$,

$$\gamma\left(\frac{\partial v'}{\partial z} - \frac{\partial w'}{\partial y}\right) = -\frac{\partial b}{\partial y}. \tag{D9}$$

The source of along-roll vorticity is the cross-wind buoyancy gradient. Substituting the first-mode expressions into Equation (D9) and matching coefficients gives

$$\gamma\left(m\,V + \frac{\mathrm{d}W}{\mathrm{d}y}\right) = \frac{\mathrm{d}B}{\mathrm{d}y}. \tag{D10}$$

One further $y$ derivative, together with (D6), produces the real-space equation

$$\frac{\mathrm{d}^2W}{\mathrm{d}y^2} - m^2W = \frac{1}{\gamma}\frac{\mathrm{d}^2B}{\mathrm{d}y^2}. \tag{D11}$$

$V$ is required to vanish far from the band, which fixes the constant of integration. Uniform buoyancy has no cross-wind gradient, so the horizontal mean of $W$ is zero.

Hats denote Fourier transforms in $y$, and $T$ is the dimensionless transfer function. Equation (D11) is solved by the multiplier

$$\widehat{W}(l) = \frac{\widehat{B}(l)}{\gamma}\,T(l), \qquad T(l) = \frac{l^2}{l^2 + m^2}, \tag{D12}$$

Here $l$ is crosswind wavenumber and $\lambda$ its wavelength, with $l = \frac{2\pi}{\lambda}$ and the $l = 0$ component set to zero. Then $\widehat{V} = \mathrm{i}\frac{m}{l}\widehat{W}$. All calculations below use this full solution, with the parameters and diagnostics in Section D.2.

When horizontal variations occur on scales much longer than the layer depth, the second-derivative term in (D11) is small relative to the other term on the left-hand side, giving

$$W \approx -\frac{1}{\gamma m^2}\frac{\mathrm{d}^2B}{\mathrm{d}y^2}. \tag{D13}$$

A warm crest then has $B'' < 0$ and $W > 0$. This relation gives a local interpretation in terms of buoyancy curvature. The relevant horizontal scale is the scale over which the profile varies, including its edge transition.

## D.2 Experimental configuration and diagnostics

WF denotes the 15 MW wind-farm simulation and CTL the no-farm simulation, with the same grid and initial and boundary conditions. The reference values are $\theta_0 = 304\,K$ and $g = 9.81\,m\,s^{-2}$.

The roll depth $H$ is estimated using the model boundary-layer depth $h_{PBL}$, defined as the height at which virtual potential temperature first exceeds its minimum in the lowest $200\,m$ by $1.25\,K$. The downstream value is the median over $x = 25 - 80\,km$ and $|y| \leq 15\,km$. At $249\,min$, $H = 1.17\,km$.

Using the friction velocity as the turbulent velocity scale and the layer depth as the mixing length, we estimate the frictional frequency as

$$\gamma = \frac{u_*}{H}, \qquad u_* = 0.68\,\mathrm{m\,s^{-1}}, \tag{D14}$$

so $\frac{1}{\gamma} \approx 29$ min. This $u_*$ is the mean surface friction velocity over this window.

For the spatial comparison, the model forcing $B(y)$ is taken from the downstream x-averaged diagnostic amplitude $B_1(y)$ at the same time, after subtraction of the mean at $|y| \geq 25\,km$. The reconstructed and simulated circulations are compared over $|y| \leq 15\,km$ and $0 \leq z \leq H$.

For the amplitude comparison, the steady response is recomputed at each output time from 150 to 276 min using the contemporaneous buoyancy profile, boundary-layer depth, and surface friction velocity. Its maximum is compared with the simulated maximum and the magnitude of the simulated minimum.

## D.3 Idealized warm-band experiment

The solution is evaluated for a band whose central virtual-temperature anomaly is the measured downstream peak, $A = 0.0168$ K. Smoothing a top-hat of width $D$ by a Gaussian of standard deviation $\delta$ gives

$$\Delta\theta_v(y) = \frac{A}{2}\left[\mathrm{erf}\left(\frac{y + D/2}{\delta\sqrt{2}}\right) - \mathrm{erf}\left(\frac{y - D/2}{\delta\sqrt{2}}\right)\right], \tag{D15}$$

and $B = \frac{g}{\theta_0} \Delta\theta_v$. The unsmoothed case is the top-hat. The standard case uses $D$=10 km, $\delta$=1 km, and the layer depth diagnosed in Section D.2.

The standard response consists of paired edge circulations (Figure 19a). Ascent lies on the inner flank of each edge and descent just outside it. The cells fill the layer; near the surface the cross-wind flow is toward the band, and it reverses at $z = H$.

For a sharp-edged band, the vertical velocity decays away from each edge over a scale of $\frac{H}{\pi}$, with little vertical motion in the band interior (Figure 19b). Smoothing broadens and weakens the response. This sensitivity suggests that resolution may affect the roll strength in CM1, motivating higher-resolution simulations in future work.

Narrowing the band moves the ascending branches inward. Their responses overlap in the narrower case, producing ascent in the interior (Figure 19c). Deepening the layer at fixed $u_*$ strengthens the circulation, because both $\gamma = \frac{u_*}{H}$ and $m = \frac{\pi}{H}$ decrease (Figure 19d).

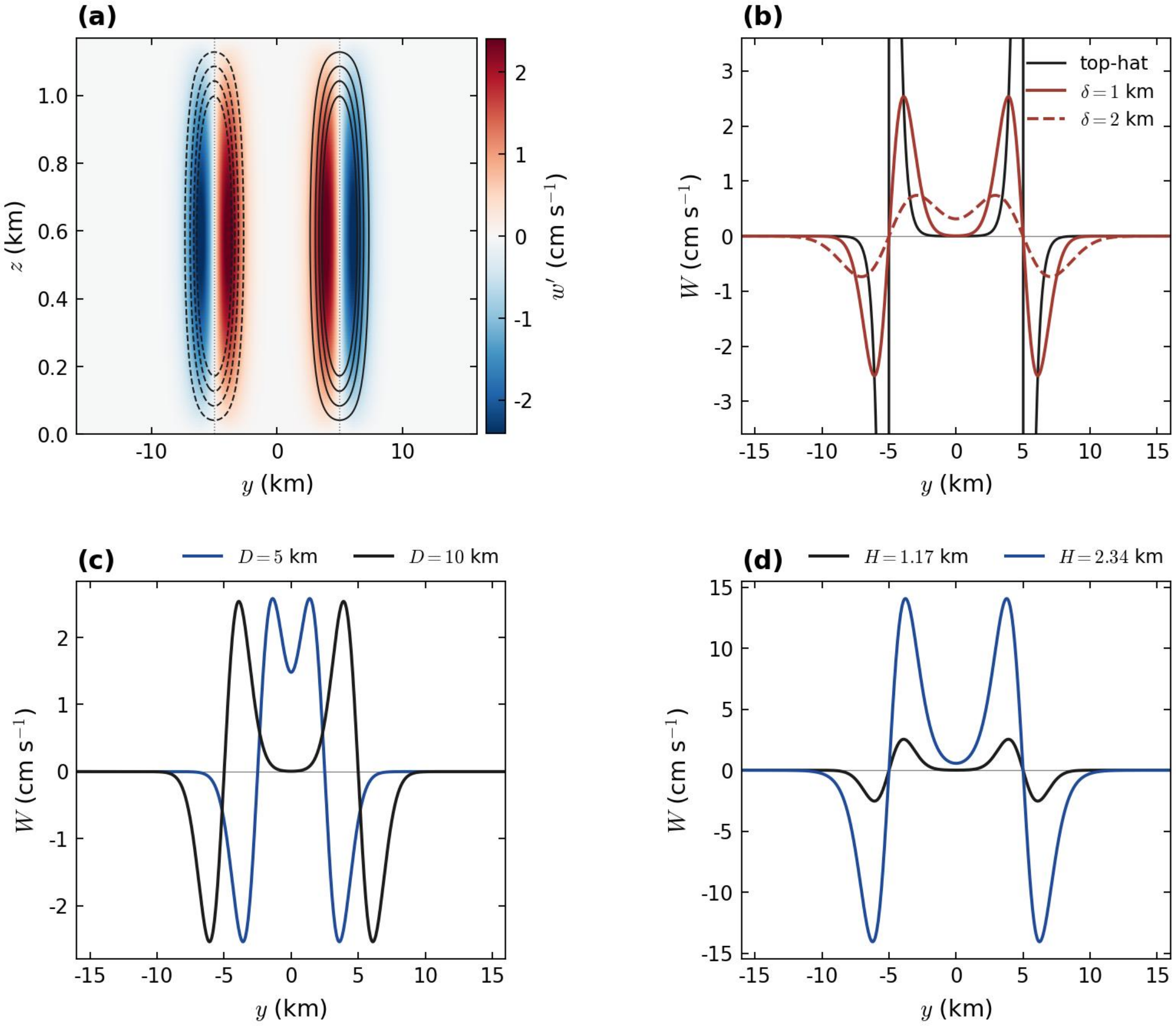

**Figure 19** Steady response of an idealized warm band whose central $\Delta\theta_v$ is $0.0168$ K, the measured downstream peak at **249 min**. Unless varied, $D = 10$ km, $\delta = 1$ km, $H = 1.17$ km. (a) Color is $w' = W\sin(mz)$.

Contours of the streamfunction satisfy $\partial\psi/\partial y = w'$; solid and dashed contours have opposite sign. Dotted lines mark the band edges $y = \pm D/2$. (b) Edge-smoothing sensitivity for a top-hat and for $\delta = 1$ and $2\ \mathrm{km}$. The top-hat is drawn on these axes and is clipped at the frame; its edge jets reach $\pm 46\ \mathrm{cm\,s^{-1}}$. (c) Band-width sensitivity for $D = 5$ and $10\ \mathrm{km}$. (d) Layer-depth sensitivity for the standard and doubled depths, with $u_* = 0.68\ \mathrm{m\,s^{-1}}$ fixed.

## D.4 Comparison of downstream circulation structure

The diagnosed downstream buoyancy profile differs from the idealized flat-topped band: it has two warm shoulders separated by a central depression (Figure 20a). In the $y$–$z$ plane, the warm anomaly extends over the farm axis through the layer below $H$ and weakens above it (Figure 20d).

The steady solution of equation (D11) for this profile places ascent near the warm shoulders and descent near the center and outside the band. The simulated $W_1$ has ascent and descent peaks that closely align with the prediction (Figure 20b). The crosswind-velocity extrema also closely align (Figure 20c). The reconstructed and simulated circulations show the same crosswind arrangement of the main cells (Figure 20e, 20f).

The simulated rolls in Figure 20f are shallower than those in Figure 20e, suggesting that the PBL-height diagnostic overestimates the effective roll depth in this case. This depth overestimate may partly explain why the steady $W$ and $V$ are stronger than the corresponding simulated first-mode amplitudes, as shown in Figure 20b, 20c and Figure 21. Incomplete adjustment toward the steady balance may also contribute, as discussed in Section D.5.

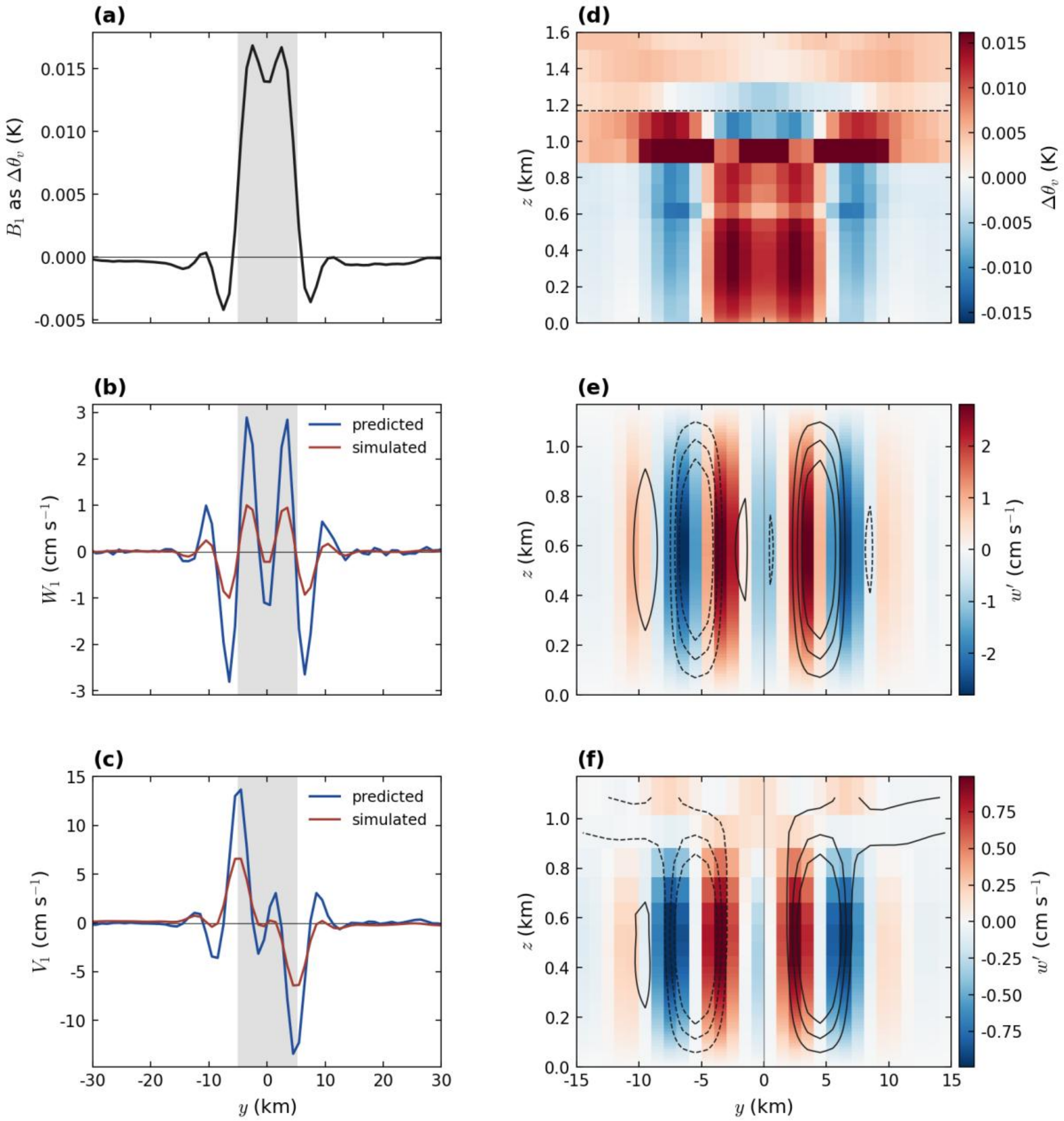

**Figure 20** Downstream structure at $\mathbf{249\,min}$, averaged over $\boldsymbol{x = 25}$–$\mathbf{80\,km}$. (a) Measured $\boldsymbol{B_1}$, drawn as $\boldsymbol{\Delta\theta_v = B_1\theta_0/g}$, after subtraction of the mean at $\boldsymbol{|y| \geq 25\,\mathrm{km}}$. Shading marks $\boldsymbol{|y| \leq 5\,\mathrm{km}}$. (b) $\boldsymbol{W}$ from (D11) applied to that profile (blue) and the simulated $\boldsymbol{W_1}$ (red). (c) $\boldsymbol{V}$ from (D6) (blue) and the simulated $\boldsymbol{V_1}$ (red). (d) Simulated $\boldsymbol{\Delta\theta_v}$ for $\boldsymbol{|y| \leq 15\,\mathrm{km}}$. The dashed line is $\boldsymbol{z = H}$. (e) Reconstructed $\boldsymbol{w' = W\sin(mz)}$ and streamfunction for $\boldsymbol{|y| \leq 15\,\mathrm{km}}$ and $\boldsymbol{0 \leq z \leq H}$. Contours satisfy $\boldsymbol{\partial\psi/\partial y = w'}$; at each height the mean of the two lateral endpoints is removed. Solid and dashed contours have opposite sign. (f) Simulated $\boldsymbol{w'}$ on the same $\boldsymbol{|y| \leq 15\,\mathrm{km}}$, $\boldsymbol{0 \leq z \leq H}$ window, with contours of $\boldsymbol{\int w'\,dy}$ drawn in the same way and a separate color scale. The vertical line in (e) and (f) is $\boldsymbol{y = 0}$.

## D.5 Comparison of circulation amplitude

Figure 21 compares the corresponding amplitude evolution. The steady solution gives the equilibrium response to the forcing at each output time. The simulated maximum remains below it, by about a factor of five at $264\,\mathrm{min}$. The amplitude difference is consistent with incomplete adjustment on the damping time scale $1/\gamma$.

After 264 min the equilibrium maximum decreases, while the simulated ascent maximum and descent magnitude continue to increase. The steady model describes the main spatial organization more closely than the amplitude evolution.

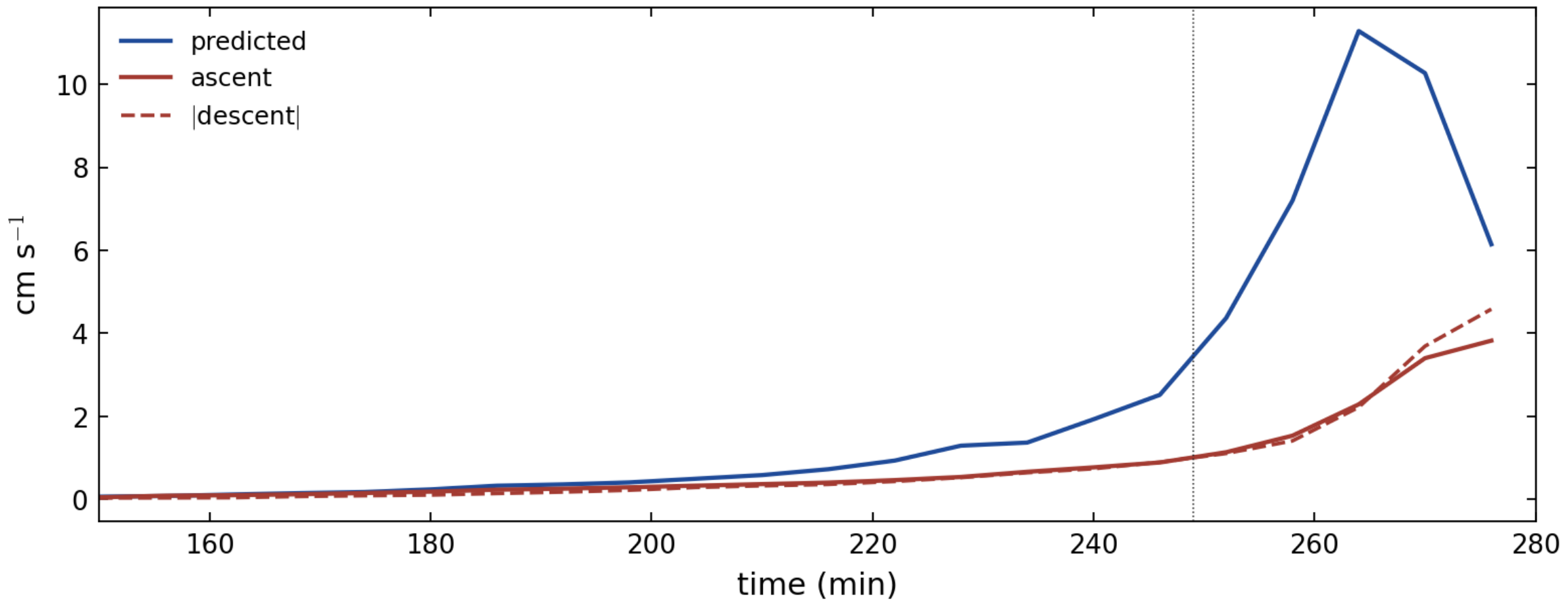

**Figure 21** Downstream vertical velocity for x=25–80 km, from 150 to 276 min. The dotted line is 249 min, the time of Figure 20. Blue: maximum of the steady $\boldsymbol{W}$ computed from $\boldsymbol{B_1(y)}$ at that time, with $\boldsymbol{H = h_{PBL}}$ and $\boldsymbol{\gamma = \frac{u_*}{H}}$. Red: maximum of the simulated $\boldsymbol{W_1}$. Red dashed: absolute value of its minimum.

## Appendix E: Sensitivity of CI to Initial Perturbations

To assess the sensitivity of CI to the initial perturbation, an additional set of simulations is conducted without the imposed potential temperature perturbation. Because CI develops later in these simulations, the domain is extended to 400 km × 100 km and the integration time to 8 h.

Wind farm related CI still develops, but generally at later times and farther downstream than in the primary experiments (Figure 22). A clear dependence on turbine rated power is retained: downstream CI occurs in all wind farm cases, but becomes progressively later and weaker as turbine rated power decreases. Convection is also more spatially organized downstream. The widespread, randomly located cells that develop at later times in the primary experiments are largely absent.

In contrast, no CI occurs in the no wind farm simulation during the full integration period. This contrast, obtained without imposed perturbations, supports the interpretation that the downstream CI analyzed in the main text is associated with the wind farm.

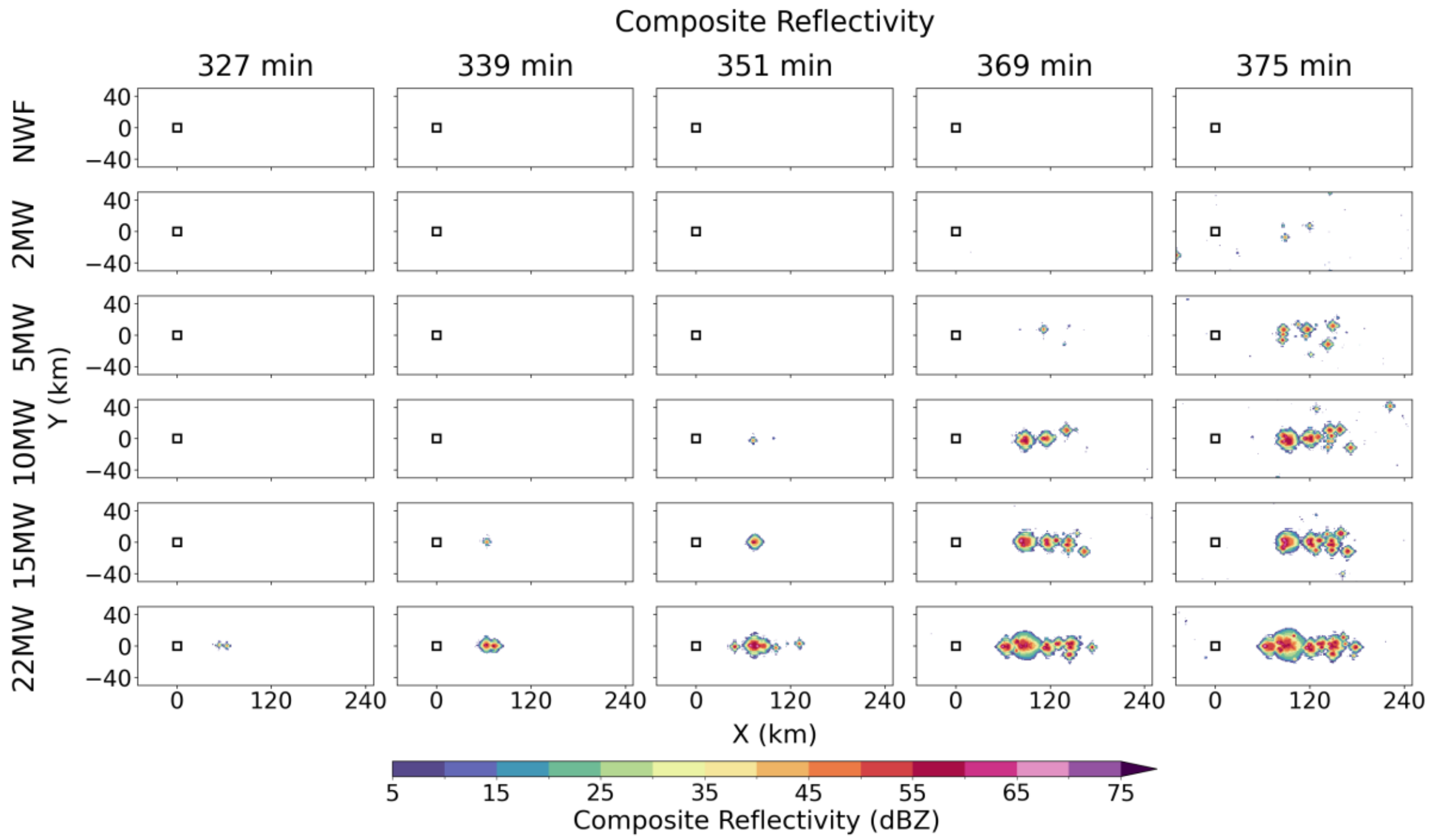

**Figure 22** Composite reflectivity from the simulations without imposed initial temperature perturbations. Composite reflectivity for the NWF and 2, 5, 10, 15, and 22 MW cases from 327 to 375 min. The black box indicates the nominal wind-farm location.

## Conflict of Interest

The authors declare there are no conflicts of interest for this manuscript.

## Acknowledgements

This work was supported by the National Training Program of Innovation for Undergraduates.

## Open Research

The CM1 model source code is publicly available from https://github.com/NCAR/CM1. The wind farm parameterization code developed in this study will be made available upon request to the corresponding author and may be considered for inclusion in a future public release of CM1. The WRF model source code is available from https://github.com/wrf-model/WRF. The NREL FLORIS software is available from https://github.com/NREL/floris. No observational data were produced in this study; all results are based on numerical simulations with configurations described in Section 2.3.

## References

Agarwal, N. J., Lundquist, J. K., Juliano, T. W., & Rybchuk, A. (2026). A North Sea in situ evaluation of the Fitch wind farm parameterization within the Mellor–Yamada–Nakanishi–Niino and 3D planetary boundary layer schemes. *Wind Energ. Sci.*, *11*(7), 2369-2403. https://doi.org/10.5194/wes-11-2369-2026

Al Fahel, N., & Archer, C. L. (2020). Observed onshore precipitation changes after the installation of offshore wind farms. *Bulletin of Atmospheric Science and Technology*, *1*(2), 179-203. https://doi.org/10.1007/s42865-020-00012-7

Ali, K., Schultz, D. M., Stallard, T., & Ouro, P. (2026). A Sensor-Driven Wind-Farm Parameterization in WRF Using a Skewed Gaussian Wind-Turbine Wake Model. *Monthly Weather Review*, *154*(6), 1155-1169. https://doi.org/10.1175/mwr-d-25-0170.1

Allaerts, D., & Meyers, J. (2017). Boundary-layer development and gravity waves in conventionally neutral wind farms. *Journal of Fluid Mechanics*, *814*, 95-130. https://doi.org/10.1017/jfm.2017.11

Angevine, W. M., Olson, J., Gristey, J. J., Glenn, I., Feingold, G., & Turner, D. D. (2020). Scale Awareness, Resolved Circulations, and Practical Limits in the MYNN–EDMF Boundary Layer and Shallow Cumulus Scheme. *Monthly Weather Review*, *148*(11), 4629-4639. https://doi.org/10.1175/mwr-d-20-0066.1

Archer, C. L., Wu, S., Ma, Y., & Jiménez, P. A. (2020). Two Corrections for Turbulent Kinetic Energy Generated by Wind Farms in the WRF Model. *Monthly Weather Review*, *148*(12), 4823-4835. https://doi.org/10.1175/mwr-d-20-0097.1

Baidya Roy, S., & Traiteur, J. J. (2010). Impacts of wind farms on surface air temperatures. *Proc Natl Acad Sci U S A*, *107*(42), 17899-17904. https://doi.org/10.1073/pnas.1000493107

Bodini, N., Lundquist, J. K., & Moriarty, P. (2021). Wind plants can impact long-term local atmospheric conditions. *Sci Rep*, *11*(1), 22939. https://doi.org/10.1038/s41598-021-02089-2

Boyer, C. H., & Keeler, J. M. (2022). Evaluation and Improvement of an Inflow-Nudging Technique for Idealized Simulations of Convective Boundary Layers. *Journal of Applied Meteorology and Climatology*, *61*(12), 1843-1860. https://doi.org/10.1175/jamc-d-22-0017.1

Bryan, G. H., & Fritsch, J. M. (2002). A Benchmark Simulation for Moist Nonhydrostatic Numerical Models. *Monthly Weather Review*, *130*, 2917-2928.

Chang, R., Sengers, B. A. M., Dörenkämper, M., & Gao, Z. (2026). Modifications to a mesoscale wind farm parameterization enhance high-altitude wind farm simulations under real-world atmospheric conditions. *Renewable Energy*, *258*. https://doi.org/10.1016/j.renene.2025.125040

Dong, L., Yu, Y., Gao, X., Zhao, S., Li, J., & Zuo, H. (2026). Potential impacts of large-scale wind farms on regional climate and vegetation in northern China: Insights from modeling studies. *Energy*, *347*. https://doi.org/10.1016/j.energy.2026.140360

Du, B., Ge, M., Li, X., & Liu, Y. (2025). A Meso-Microscale Coupled Wind Farm Parameterization. *Boundary-Layer Meteorology*, *191*(8). https://doi.org/10.1007/s10546-025-00928-7

Du, B., Li, Q., Ge, M., Li, X., & Liu, Y. (2026). Gaussian-based multicolumn spatial distribution method for wind farm parameterizations. *Physical Review Fluids*, *11*(2). https://doi.org/10.1103/5lsj-m746

Durran, D. R., & Klemp, J. B. (1983). A Compressible Model for the Simulation of Moist Mountain Waves. *Monthly Weather Review*, *111*(12), 2341–2361. https://doi.org/10.1175/1520-0493(1983)111<2341:ACMFTS>2.0.CO;2

Englberger, A., & Dörnbrack, A. (2017). Impact of the Diurnal Cycle of the Atmospheric Boundary Layer on Wind-Turbine Wakes: A Numerical Modelling Study. *Boundary-Layer Meteorology*, *166*(3), 423-448. https://doi.org/10.1007/s10546-017-0309-3

Fiedler, B. H., & Bukovsky, M. S. (2011). The effect of a giant wind farm on precipitation in a regional climate model. *Environmental Research Letters*, *6*(4). https://doi.org/10.1088/1748-9326/6/4/045101

Fischereit, J., Brown, R., Larsén, X. G., Badger, J., & Hawkes, G. (2021). Review of Mesoscale Wind-Farm Parametrizations and Their Applications. *Boundary-Layer Meteorology*, *182*(2), 175-224. https://doi.org/10.1007/s10546-021-00652-y

Fitch, A. C., Barstad, I., Michalakes, J., Gupta, A. K., Dudhia, J., Lundquist, J. K., & Olson, J. B. (2012). Local and Mesoscale Impacts of Wind Farms as Parameterized in a Mesoscale NWP Model. *Monthly Weather Review*, *140*(9), 3017-3038. https://doi.org/10.1175/mwr-d-11-00352.1

Fitch, A. C., Lundquist, J. K., & Olson, J. B. (2013). Mesoscale Influences of Wind Farms throughout a Diurnal Cycle. *Monthly Weather Review*, *141*(7), 2173-2198. https://doi.org/10.1175/mwr-d-12-00185.1

Fitch, A. C., Olson, J. B., & Lundquist, J. K. (2013). Parameterization of Wind Farms in Climate Models. *Journal of Climate*, *26*(17), 6439-6458. https://doi.org/10.1175/jcli-d-12-00376.1

Gadde, S. N., & Stevens, R. J. A. M. (2019). Effect of Coriolis force on a wind farm wake. *Journal of Physics: Conference Series*, *1256*, 012026. https://doi.org/10.1088/1742-6596/1256/1/012026

García-Santiago, O., Hahmann, A. N., Badger, J., & Peña, A. (2024). Evaluation of wind farm parameterizations in the WRF model under different atmospheric stability conditions with high-resolution wake simulations. *Wind Energy Science*, *9*(4), 963-979. https://doi.org/10.5194/wes-9-963-2024

GWEC. (2025). *Global Wind Report 2025*. G. W. E. Council.

GWEC. (2026). *Global Wind Report 2026*. G. W. E. Council.

Jia, Z., Yang, X., Chen, A., Yang, D., Zhang, M., & Wei, L. (2024). Localized Eco-Climatic Impacts of Onshore Wind Farms: A Review. *Journal of Resources and Ecology*, *15*(1). https://doi.org/10.5814/j.issn.1674-764x.2024.01.013

Jiang, H., Xie, M., Li, X., Tian, M., Cui, W., & Hao, D. (2025). Wind Farms Impacts on Land Surface Temperature and Its Driving Factors in an Arid Area of Xinjiang, China. *Sustainability*, *17*(21). https://doi.org/10.3390/su17219445

Klemp, J. B., & Wilhelmson, R. B. (1978). The Simulation of Three-Dimensional Convective Storm Dynamics. *Journal of the Atmospheric Sciences*, *35*(6), 1070–1096. https://doi.org/10.1175/1520-0469(1978)035<1070:TSOTDC>2.0.CO;2

Li, Z., Liu, X., & Yang, X. (2022). Review of Turbine Parameterization Models for Large-Eddy Simulation of Wind Turbine Wakes. *Energies*, *15*(18). https://doi.org/10.3390/en15186533

Liu, N., Zhao, X., Zhang, X., Zhao, J., Wang, H., & Wu, D. (2023). Remotely sensed evidence of the divergent climate impacts of wind farms on croplands and grasslands. *Science of The Total Environment*, *905*. https://doi.org/10.1016/j.scitotenv.2023.167203

Ma, Y., Archer, C. L., & Vasel-Be-Hagh, A. (2022). The Jensen wind farm parameterization. *Wind Energy Science*, *7*(6), 2407-2431. https://doi.org/10.5194/wes-7-2407-2022
Mantovani, J. A., Carneiro, R., Borges, C. K., Ibarra-Espinosa, S., Aravéquia, J. A., Fisch, G., & Herdies, D. L. (2026). Impact of PBL Schemes on the Simulation of PBL Height in the Central Amazon Basin. *Geosciences*, *16*(4). https://doi.org/10.3390/geosciences16040134
Mecikalski, J. R., & Bedka, K. M. (2006). Forecasting Convective Initiation by Monitoring the Evolution of Moving Cumulus in Daytime GOES Imagery. *Monthly Weather Review*, *134*(1), 49-78. https://doi.org/10.1175/MWR3062.1
Meng, Z., Cui, Q., Wang, C., Liu, H., Zheng, Z., & Wang, X. (2026). *Impact of Desert Wind Farms on Convection Initiation* EGU General Assembly 2026, Vienna, Austria.
Olson, J. B., Angevine, W. M., Jensen, A., Grell, G., Smirnova, T., & Toy, M. (2024). *Development Strategy for GSL-led to Physical Parameterization for all Model Hosts/Applications* [Conference Presentation].
Onwukwe, C., Jackson, P. L., & Déry, S. J. (2022). Intercomparison of atmospheric forcing datasets and two PBL schemes for precipitation modelling over a coastal valley in northern British Columbia, Canada. *Meteorological Applications*, *29*(1). https://doi.org/10.1002/met.2043
Orlanski, I. (1976). A simple boundary condition for unbounded hyperbolic flows. *Journal of Computational Physics*, *21*(3), 251-269. https://doi.org/10.1016/0021-9991(76)90023-1
Peña, A., Mirocha, J. D., & van der Laan, M. P. (2022). Evaluation of the Fitch Wind-Farm Wake Parameterization with Large-Eddy Simulations of Wakes Using the Weather Research and Forecasting Model. *Monthly Weather Review*, *150*(11), 3051-3064. https://doi.org/10.1175/mwr-d-22-0118.1
Porté-Agel, F., Bastankhah, M., & Shamsoddin, S. (2020). Wind-Turbine and Wind-Farm Flows: A Review. *Boundary Layer Meteorol*, *174*(1), 1-59. https://doi.org/10.1007/s10546-019-00473-0
Prein, A. F., Ban, N., Ou, T., Tang, J., Sakaguchi, K., Collier, E., Jayanarayanan, S., Li, L., Sobolowski, S., Chen, X., Zhou, X., Lai, H.-W., Sugimoto, S., Zou, L., Hasson, S. u., Ekstrom, M., Pothapakula, P. K., Ahrens, B., Stuart, R.,…Chen, D. (2022). Towards Ensemble-Based Kilometer-Scale Climate Simulations over the Third Pole Region. *Climate Dynamics*, *60*(11-12), 4055-4081. https://doi.org/10.1007/s00382-022-06543-3
Quint, D., Lundquist, J. K., & Rosencrans, D. (2025). Simulations suggest offshore wind farms modify low-level jets. *Wind Energy Science*, *10*(1), 117-142. https://doi.org/10.5194/wes-10-117-2025
Rajewski, D. A., Takle, E. S., Lundquist, J. K., Oncley, S., Prueger, J. H., Horst, T. W., Rhodes, M. E., Pfeiffer, R., Hatfield, J. L., Spoth, K. K., & Doorenbos, R. K. (2013). Crop Wind Energy Experiment (CWEX): Observations of Surface-Layer, Boundary Layer, and Mesoscale Interactions with a Wind Farm. *Bulletin of the American Meteorological Society*, *94*(5), 655-672. https://doi.org/10.1175/bams-d-11-00240.1
Raymond, W. H., & Kuo, H. L. (1984). A radiation boundary condition for multi-dimensional flows. *Quarterly Journal of the Royal Meteorological Society*, *110*(464), 535-551. https://doi.org/10.1002/qj.49711046414
Redfern, S., Olson, J. B., Lundquist, J. K., & Clack, C. T. M. (2019). Incorporation of the Rotor-Equivalent Wind Speed into the Weather Research and Forecasting Model's

Wind Farm Parameterization. *Monthly Weather Review*, *147*(3), 1029-1046. https://doi.org/10.1175/mwr-d-18-0194.1

Smith, R. B. (2010). Gravity wave effects on wind farm efficiency. *Wind Energy*, *13*(5), 449-458. https://doi.org/10.1002/we.366

Su, N., Li, X., Lyu, X., Dang, D., Liu, S., & Zhang, C. (2025). Impact pathways of wind farms on grassland carbon and water cycles. *Journal of Environmental Management*, *388*. https://doi.org/10.1016/j.jenvman.2025.126036

Tomaszewski, J. M., & Lundquist, J. K. (2021). Observations and simulations of a wind farm modifying a thunderstorm outflow boundary. *Wind Energy Science*, *6*(1), 1-13. https://doi.org/10.5194/wes-6-1-2021

van der Laan, M. P., & Sørensen, N. N. (2017). Why the Coriolis force turns a wind farm wake clockwise in the Northern Hemisphere. *Wind Energ. Sci.*, *2*(1), 285-294. https://doi.org/10.5194/wes-2-285-2017

Volker, P. J. H., Badger, J., Hahmann, A. N., & Ott, S. (2015). The Explicit Wake Parametrisation V1.0: a wind farm parametrisation in the mesoscale model WRF. *Geoscientific Model Development*, *8*(11), 3715-3731. https://doi.org/10.5194/gmd-8-3715-2015

Vollmer, L., Sengers, B. A. M., & Dörenkämper, M. (2024). Brief communication: A simple axial induction modification to the Weather Research and Forecasting Fitch wind farm parameterization. *Wind Energy Science*, *9*(8), 1689-1693. https://doi.org/10.5194/wes-9-1689-2024

Warner, T. T., Peterson, R. A., & Treadon, R. E. (1997). A Tutorial on Lateral Boundary Conditions as a Basic and Potentially Serious Limitation to Regional Numerical Weather Prediction. *Bulletin of the American Meteorological Society*, *78*(11), 2599–2618. https://doi.org/10.1175/1520-0477(1997)078<2599:ATOLBC>2.0.CO;2

Weisman, M. L., & Klemp, J. B. (1982). The Dependence of Numerically Simulated Convective Storms on Vertical Wind Shear and Buoyancy. *Monthly Weather Review*, *110*(6), 504-520. https://doi.org/10.1175/1520-0493(1982)110<0504:TDONSC>2.0.CO;2

Wu, S., & Archer, C. L. (2021). Near-Ground Effects of Wind Turbines: Observations and Physical Mechanisms. *Monthly Weather Review*, *149*(3), 879-898. https://doi.org/10.1175/mwr-d-20-0186.1

Wu, S., Archer, C. L., & Mirocha, J. D. (2024). New insights on wind turbine wakes from large-eddy simulation: Wake contraction, dual nature, and temperature effects. *Wind Energy*, *27*(11), 1130-1151. https://doi.org/10.1002/we.2827

Xia, G., Zhou, L., Minder, J. R., Fovell, R. G., & Jimenez, P. A. (2019). Simulating impacts of real-world wind farms on land surface temperature using the WRF model: physical mechanisms. *Climate Dynamics*, *53*(3-4), 1723-1739. https://doi.org/10.1007/s00382-019-04725-0

Zhou, L., Baidya Roy, S., & Xia, G. (2022). Weather, Climatic and Ecological Impacts of Onshore Wind Farms. In *Comprehensive Renewable Energy* (pp. 165-188). https://doi.org/10.1016/b978-0-12-819727-1.00018-2

Zhou, L., Tian, Y., Baidya Roy, S., Thorncroft, C., Bosart, L. F., & Hu, Y. (2012). Impacts of wind farms on land surface temperature. *Nature Climate Change*, *2*(7), 539-543. https://doi.org/10.1038/nclimate1505